\documentclass[12pt,preprint]{aastex}

\newcommand{\ea}{\it et al.}

\newcommand{\rg}{ radio ~galaxies~}
\newcommand{\qso}{quasars~}
\shorttitle{Far-IR Properties of Optical Quasars}
\shortauthors{Andreani \ea}
\begin{document}
\title{The dusty environment of \qso. Far-IR properties of Optical Quasars
\footnote{
based on observations with ISO, an ESA project with   
instruments funded by ESA Member States   
(especially the PI countries: France, Germany, the Netherlands and the   
United Kingdom) with the participation of ISAS and NASA}
}

\author{Paola Andreani\altaffilmark{1}}
\affil{INAF - Osservatorio Astronomico Padova,
Vicolo dell'Osservatorio 5, Padova, I-35122, Italy, e-mail:andreani@pd.astro.it}

\author{Stefano Cristiani\altaffilmark{2}}
\affil{INAF - Osservatorio Astronomico di Trieste,
Via G.B.Tiepolo 11 - I 34131 Trieste, Italy, e-mail:cristiani@ts.astro.it}

\author{Andrea Grazian\altaffilmark{3}}
\affil{Dipartimento di Astronomia, Universit\`a di Padova,
Vicolo dell'Osservatorio 2, Padova, I-35122, Italy}

\author{Fabio La Franca}\affil{
Dipartimento di Fisica, Universit\`a degli Studi "Roma Tre", viale della
     Vasca Navale 84, I-00146 Roma, Italy, e-mail:lafranca@fis.uniroma3.it}

\author{Pippa Goldschmidt}
\affil{Astrophysics Group, Blackett Laboratory, Imperial College of Science
Technology and Medicine, Prince Consort Road, London SW7 2BZ, UK }

\altaffiltext{1}{
Max-Planck I. f\"ur Extraterrestrische
Physik, Postfach 1312, 85741 Garching, Germany, e-mail:andreani@mpe.mpg.de}
\altaffiltext{2}{
Space Telescope - European Coordinating Facility
European Southern Observatory,
K. Schwarzschild str.2, D-85748 Garching, Germany,
e-mail:scristia@eso.org}
\altaffiltext{3}{European Southern Observatory,
K. Schwarzschild str.2, D-85748 Garching, Germany,
agrazian@eso.org}

\begin{abstract}
We present the ISO far-IR photometry of a complete sub-sample of optically
selected bright quasars belonging to two complete
surveys selected through multicolour (U,B,V,R,I) techniques.
The ISOPHOT camera on board of the ISO Satellite was used to target these
quasars at wavelengths of 7.3, 11.5, 60, 100 and 160 $\mu m$.
Almost two thirds of the objects
were detected at least in one ISOPHOT band. The detection rate is
independent of the source redshift, very likely due to
the negative K-correction of the far-IR thermal emission.
More than a half of the optically selected QSOs show significant
emission between 4 and 100 $\mu m$ in the quasar rest-frame.\hfill\break
These fluxes have a very likely thermal origin, although in a few objects
an additional contribution from a non-thermal component is plausible
in the long wavelength bands. In a colour-colour diagram these objects
span a wide range of properties from AGN-dominated to ULIRG-like.
The far-IR composite spectrum of the quasar population presents
a broad far-IR bump between 10 and 30\,$\mu$m and a sharp drop at
$\lambda >$100\,$\mu$m in the quasar restframe.
The amount of energy emitted in the far-IR,
is on average a few times larger than that emitted in the blue
and the ratio $\frac{L_{\rm FIR}}{L_{\rm B}}$ increases with the
bolometric luminosity.
Objects with fainter blue magnitudes have larger ratios between the far-IR
($\lambda > 60 \mu m$) fluxes and the blue band flux,
which is attributed to extinction by dust around the central source.
No relation between the blue absolute magnitude and the
dust colour temperature is seen, suggesting that 
the dominant source of FIR energy could be linked to a concurrent starburst
rather than to gravitational energy produced by the central engine.
\end{abstract}



\section{Introduction}

The properties of the circumnuclear environment are a key element
to understand the formation and evolution of AGN. In particular, 
while these properties are well studied in the UV/optical spectral domain
their mid-IR and far-IR characteristics are still a matter of debate. 

The sampling of the far-IR part of the Spectral Energy Distribution (SED)
is important for determining not only the overall energy balance but also
to discriminate the dominant emitting mechanism and to trace any evolutionary
path from a dust-enshrouded phase to a dust-free {\it regime}, corresponding to
a "neat" optical phase (Sanders et al., 1989).
Possibly related is the investigation of the structure and anisotropy
of the obscuring-reprocessing torus as predicted by unification models of AGN.
In particular, knowledge of the spectral region where the dust emission peaks 
and starts to become optically thick is needed to constrain the nature of
the torus and its fundamental 
geometrical and physical parameters (dimensions, dust temperature and optical 
depth) providing an
orientation-independent parameter to test the unification hypothesis.

IRAS observations have offered the first glimpse to the far-IR
properties of optical quasars
but did not have the required sensitivity to yield an unbiased view
of the AGN class as a whole. In fact, only known optical quasars (PG quasars),
with a very narrow range of fluxes and consequently a narrow range of
luminosities at any given redshift, have been
targetted and studied. The lack of any statistically well-defined sample
prevented from any inference on the general FIR properties 
of {\it optically-selected quasars}.
Also subsequent studies, based on {\it mm} emission of IRAS
QSOs and of high-z optical QSOs (Chini et al., 1989, Omont et al. 1996,
Andreani et al., 1999, Carilli et al., 2000, Omont et al., 2001,
and references therein) and on ISO observations
(Haas et al., 1998, 2000; Polletta et al., 2000;
Oyabu et al., 2001), could not address this issue and
mainly focussed on the question of the
emitting mechanism.

The sampling of the shortest and longest parts of the IR spectrum
via ISO observations is a key test to disentangle the emitting mechanism.
However a typical conclusion of ISO studies is that it is not easy to
disentangle pure AGN emission from a likely common FIR-emitting
star-formation component. There is not a general agreement among
different observations (Haas et al., 1998; 2000; Polletta et al., 2000)
and interpreting models (e.g. Granato \& Danese, 1994; Rowan-Robinson, 1995).

In this paper we report ISO observations of a complete
optically selected sample of bright (15$\leq m_B\leq$17) quasars.
The original aim of our programme was to use these data to infer any
evolutionary link from "pure" AGN objects to eventual transition objects
linking AGN to ULIRGs. The investigation of the complete sample would have
allowed us to perform a number of statistical tests
(for instance the evaluation
of the bivariate luminosity function at optical-IR wavelengths)
to study how the optical-infrared relation for quasars
changes as a function of both luminosity and redshift.
Because of the reduced sensitivity of ISOPHOT the number of observed objects
is greatly reduced from the original one and the number of detection
is also consequently too small to allow reliable statistics. 
Neverthless, a number of different properties of the far-IR emission of
optical quasars can be inferred from these data and help understanding
the expected behaviour of the type-1 AGN population in surveys from
future space mission -- such as SIRT-F and HSO.


\section{The Sample}
A statistically well defined sample of bright quasars was selected from
two complete optical multicolour (U,B,V,R,I) quasar surveys: the
Edinburgh Bright Quasar Survey (Goldschmidt et al., 1992)
and the ESO Key-Programme "A Homogeneous Bright Quasar Survey"
\cite{Cri95} in a total sky area of 888 sq. deg.
The optical fluxes are in the range 15 $<$ B $<$ 17, partially overlapping
those of the PG survey, but with a much better photometric accuracy
($\sigma_B \sim$0.1mag).
The objects finally observed with ISO are listed in Table 1, where
names (column 1), J2000 coordinates (2-3), redshifts (4),
{\bf Johnson Cousins U,B,V,R,I} magnitudes (columns 5-9), and the absolute
B magnitudes are reported.
The apparent B-magnitude, m$_{\rm B}$, were taken from the
Homogeneous Bright QSO survey (HBQS) \cite{Cri95}
whereas B$_j$ apparent magnitudes were derived from the
Digitized Sky Survey (DSS)
\footnote{
The online Digitized Sky Surveys (http://archive.eso.org/dss/dss) server at
the ESO/ST-ECF
Archive provides access to the CD-ROM set produced by the Space
Telescope Science Institute
through its Guide Star Survey group.} calibrated plates (Grazian et al.,
2001).
K-corrections were inferred from a QSO composite spectrum \cite{crivio}.
The apparent magnitudes are corrected for Galactic absorption using the
reddening maps
derived from HI and Galaxy Counts by Burstein \& Heiles (1982) 
and the luminosity distances are derived according to the adopted Cosmology
(H$_0$=50 km/s/Mpc, $\Lambda=0$, $\Omega=1$).
\section{The Observations}

\subsection{The ISOPHOT Observations}

The ISOPHOT data presented here result from the merging of two distinct
Open Time programmes PANDREAN and PGOLDSCH.
\hfill\break
The observed sample consists of 34 quasars 
of which 21 were observed at 11.5, 60 and 160 \,$\mu$m and 13
at 7.5, 25, 60 and 100 \,$\mu$m. The log of the ISOPHOT Observations
is in Table 2, {\bf coordinates used for pointing are those listed in Table 1}.

The data were reduced with the ISOPHOT Interactive Analysis Tool (PIA Version
V8.2, Gabriel et al., 1998).
All data  were corrected for electronics nonlinearity and cosmic particle
glitches.
Chopped data were then averaged: 
all the points corresponding to one chopper position were averaged and
fitted with a polynomial curve to estimate their time
behaviour. If the detector shows stability, i.e. the fitted curve converges
to a constant value, this latter is used as the signal related to that
chopper position. In some cases, when the integration time on source was too
short, for the ISOPHOT detectors to reach the stability, the signal value
was inferred by extrapolating the time-dependent signal curve to
an asimptotic constant value.
Data are further corrected for glitches, for orbital dependent dark current and for
the signal dependence on the ramp integration time.
\hfill\break
The same reduction procedure was applied to the associated measurements of
the thermal Fine Calibration Source (FCS1), which were used to estimate fluxes.

For raster maps additional softwares were used to extract the signal,
kindly provided by Ilse van Bemmel
and Martin Haas and are described in van Bemmel et al. (2000) and
Haas et al. (2000).

Table 3 lists the ISOPHOT fluxes at 7, 12, 25, 60,
100 and 160 $\mu m$ (columns 2-7).
\hfill\break
30 \% of the objects have fluxes larger than 3 $\sigma$ at 60\,$\mu$m, 
5 out of 7 were detected with raster maps while only 5 out of 27 with 
chopper mode. No objects were detected at 7 and 12 $\mu m$ (only
in chopping mode) and only 4 at 25 and 100\,$\mu$m.
At 160\,$\mu$m 6 objects were detected (3 detections obtained
from the 7 raster scannings).
The raster scanning observing mode shows the highest sensitivity. However at
160\,$\mu$m the sensitivity could be reduced in regions where
the Galactic background due to cirrus emission is high.
We checked detections against cirrus emission level and the results are
shown in the Appendix.

   \begin{figure*}
      \plotone{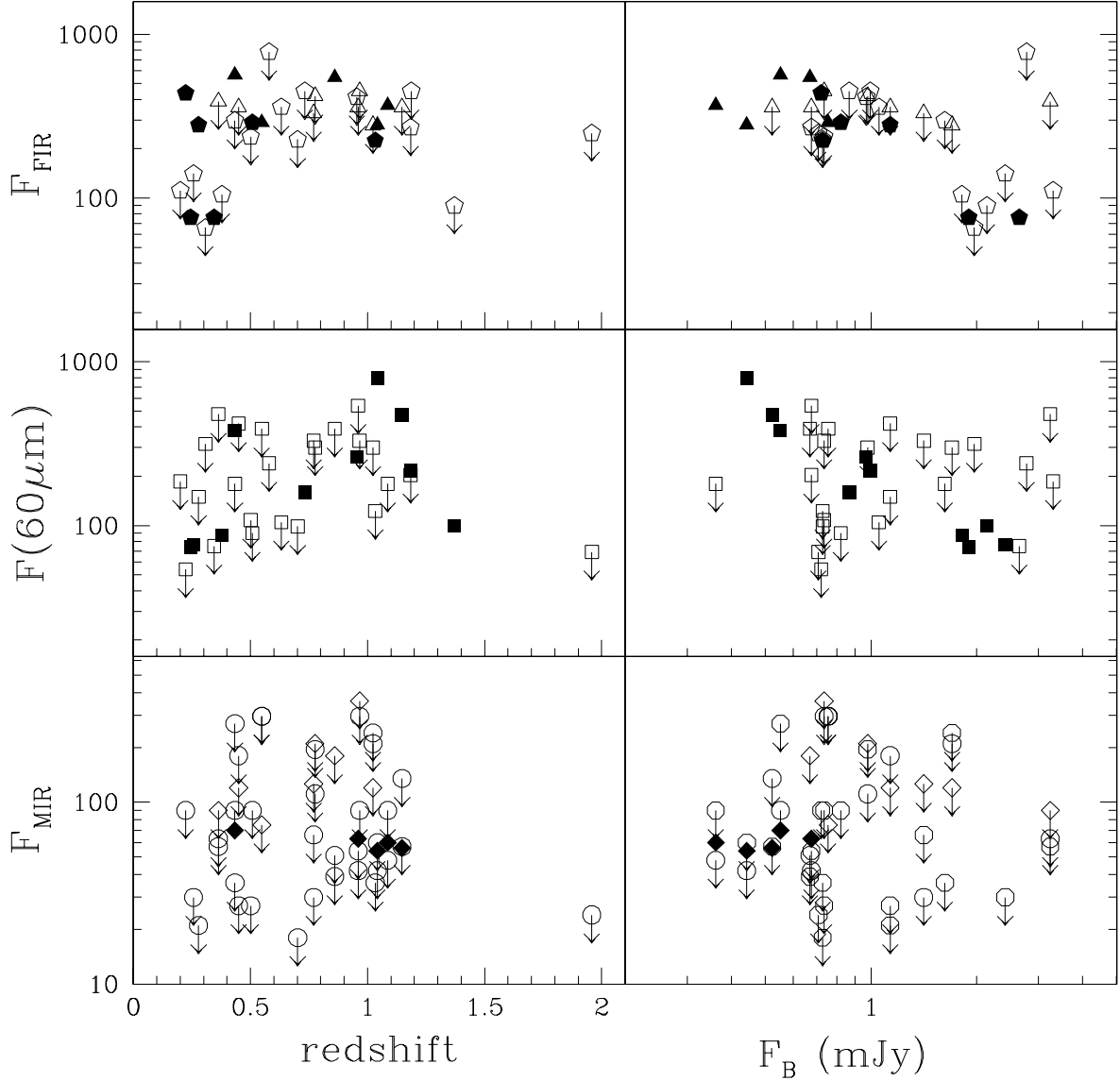}
      \caption{Logarithm of the ISOPHOT fluxes in mJy as a function of
redshift (left panels)
and of the B fluxes in mJy (right panels). The lower panels report the
fluxes at 7 and 12 \,$\mu$m (circles) and 25 \,$\mu$m (diamonds),
the upper panels the fluxes at 100 \,$\mu$m (triangles) and 160
\,$\mu$m (pentagons). Filled symbols correspond to detections, open to upper
limits. No clear dependence is seen: upper limits and
detections are equally distributed (see text). A slight
trend -- lower IR fluxes at large B fluxes
-- exists for those objects detected at 60 and 160\,$\mu$m.
         \label{fig:fluxes}}
 \end{figure*}

Figure \ref{fig:fluxes} provides a general view of all the ISOPHOT fluxes
collected in the present work. The left panels show them as a function of
redshift, the right panels as a function of the B-band flux.
{\rm MIR} is the observed wavelength
range 7 - 25 $\mu m$ and {\rm FIR} 100- 160 $\mu m$.
\hfill\break
The lack of detections, shown in Figure \ref{fig:fluxes},
can be ascribed not only to the reduced sensitivity in the chopper
mode but also to the nature of the FIR emission in optical quasars,
showing a deficit in the short (7 and 12\,$\mu$m) and long
(160\,$\mu$m) observed wavelengths emission,
because of the characteristic high temperature of the dust grains
heated by the AGN.
\hfill\break
Although many of the values reported in Figure \ref{fig:fluxes} are upper
limits, it emerges an absence of dependence of the detection rate with
redshift:
the detections are {\it randomly} spread all over the redshift range.
10 detected objects have redshifts larger than the mean redshift
($\langle z \rangle = 0.72$, median redshift is 0.7) of the
sample: $z > \langle z \rangle$,
while the other 10 detections have $z < \langle z \rangle$.
Similarly there is little correlation between the B apparent
luminosity and the FIR fluxes, with only a slight trend
of larger FIR fluxes for smaller B flux.
The most B-luminous objects are not correspondingly FIR-luminous and there is no scaling of the FIR fluxes 
according to the B-band magnitude as it was assumed computing the
integration time
(the smaller the magnitude the shorter the integration time).
The latter was computed aiming at a uniform sensitivity in terms of
$\alpha _{oIR} = \log F_{\rm blue} - \log F_{\rm IR} \sim -2 $.

Only four objects are detected in more than one IR band:
0144-39, 1404+09, 1415+00 and 1415-00.
Individual notes are given in Section 4.

\subsection{Additional data}

\subsubsection{IRAS SCANPI data}
The IRAS data listed in table 3 are the result of
co-added fluxes provided by IPAC 
SCANPI (Scan Processing and Integration Tool) program. This procedure
performs a one-dimensional coaddition of all the
IRAS survey data on the source. The sensitivity is comparable to that
achieved by the FSC (Faint Source Catalog) for point sources
(see the IPAC manual for details, http://ipac.caltech.edu/). 
\hfill\break
Tentative (3$\sigma$) detections among IRAS data are found for 0120-28,
0144-39, 1252+02, 1321+28, 1326-05, 1351+01, 1355+02, 1415+00, 1415-00,
2313-30, 2357-35 and are shown in boldface in table 3.
\hfill\break
{\bf Some objects in table 3 show discrepant 60$\mu$m fluxes between the
ISO and IRAS
measurements. Some authors (e.g. Alton et al., 1998;
Spinoglio et al., 2002) have argued that
ISO fluxes are affected by a calibration error as large as 30\%
and found ISO fluxes in excess
with respect the corresponding IRAS ones. However in some cases (1415-00
and 1415+00), even taking into
account an additional 30 \% error ISO values lie well above the IRAS ones.
As shown in Appendix A, it is also not straightforward to attribute this
discrepancy to additional cirrus emission. This contribution is expected
to be larger in IRAS measurements because of the wider IRAS beams.
This discrepancy may be due to variability but, at present,
this issue cannot be settled and could only be addressed by regular
radio observations.}

\subsubsection{near-IR Photometry}

{\bf 
Counterparts within 4$^{\prime\prime}$ of the QSO listed in Table 1
were detected using the 2MASS Second Incremental Data Release
(http://www.ipac.caltech.edu/2mass/)
and the corresponding J (1.25$\mu$m), H (1.65$\mu$m)
and K$_s$ (2.17$\mu$m) photometry of the 19 (out of 34) objects found
is listed in table 1 (see also Barkhouse \& Hall 2001).
All objects with near-IR counterparts show rising near-IR spectra with
magnitudes decreasing from the J to the K-band.
}
\subsubsection{radio data}

Radio data are used in this work to check whether strong non-thermal
emission showing up in the radio domain would possibly
affect the far-IR fluxes. The data
were taken from the NVSS (NRAO VLA Sky Survey)
survey \citep{Con98} and from the FIRST survey \citep{bec94}
and are listed in table 1 for the 21 objects with sky positions
covered by these surveys (at $\delta > -40^\circ$).
{\bf 8 objects out of 21 observed in these surveys have a logarithmic
ratio between the 1.4 GHz and the B-band luminosities,
$\frac{L_{\rm 1.4GHz}}{L_{\rm B}}$, larger than 1,
i.e. are radio-loud objects}.

   \begin{figure*}
    \plotone{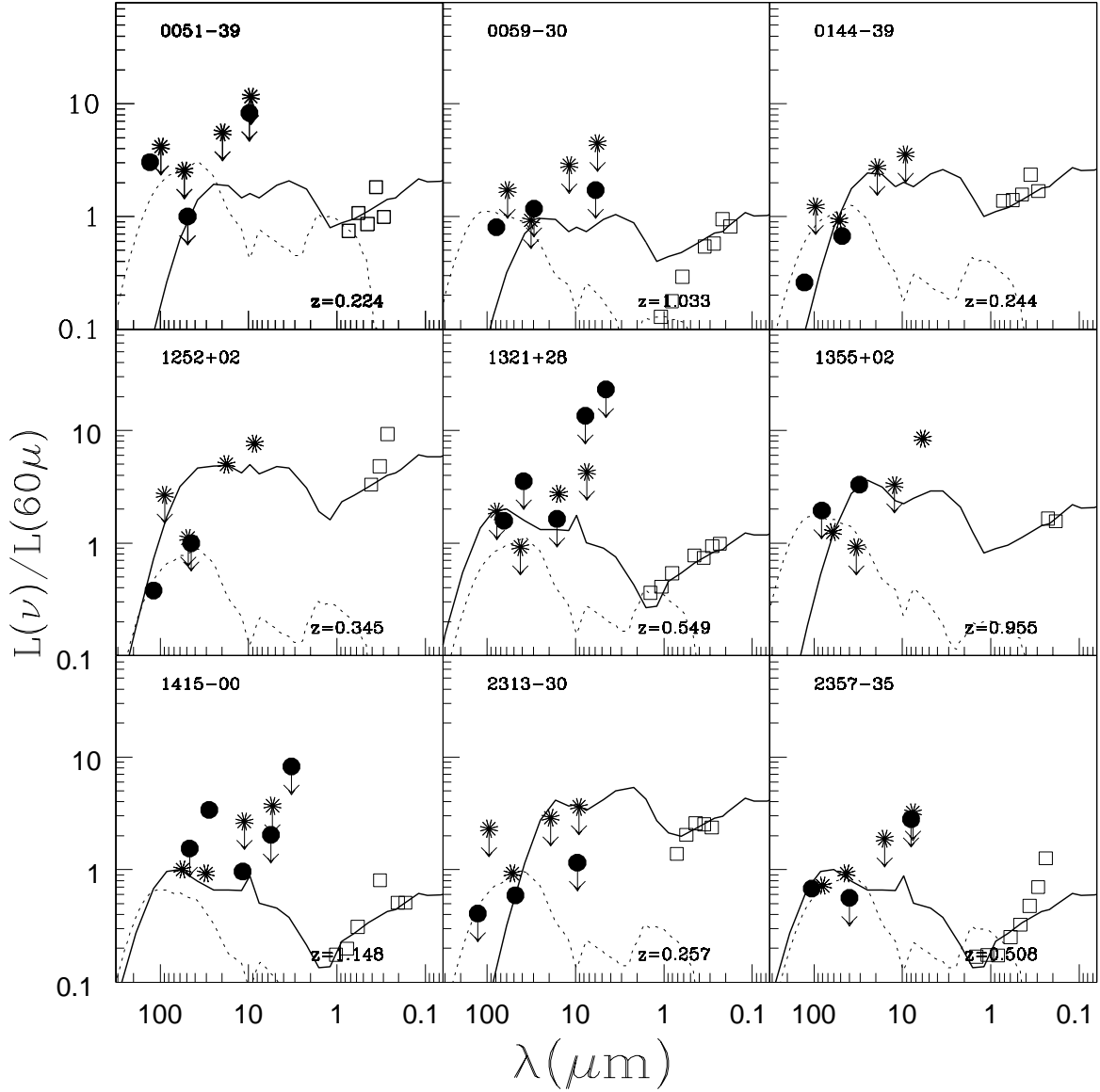}
	\caption{Spectral Energy Distributions (SEDs) 
as $\nu L(\nu)$ normalized
at 60 $\mu m$ of the objects reported in tables 1$\div$3 having
at least one ISOPHOT detection. Open squares refer to optical and near-IR
photometry, filled circles to ISOPHOT data, asterisks to IRAS data.
The solid line corresponds to one of
the models by Granato \& Danese (1994) predicting the emission of dust distributed
in the surrounding medium of the central AGN (see text). The dotted line
is the spectrum of M82 shown here as a prototype starburst.
              \label{fig:SED1}}
    \end{figure*}

   \begin{figure*}
   \plotone{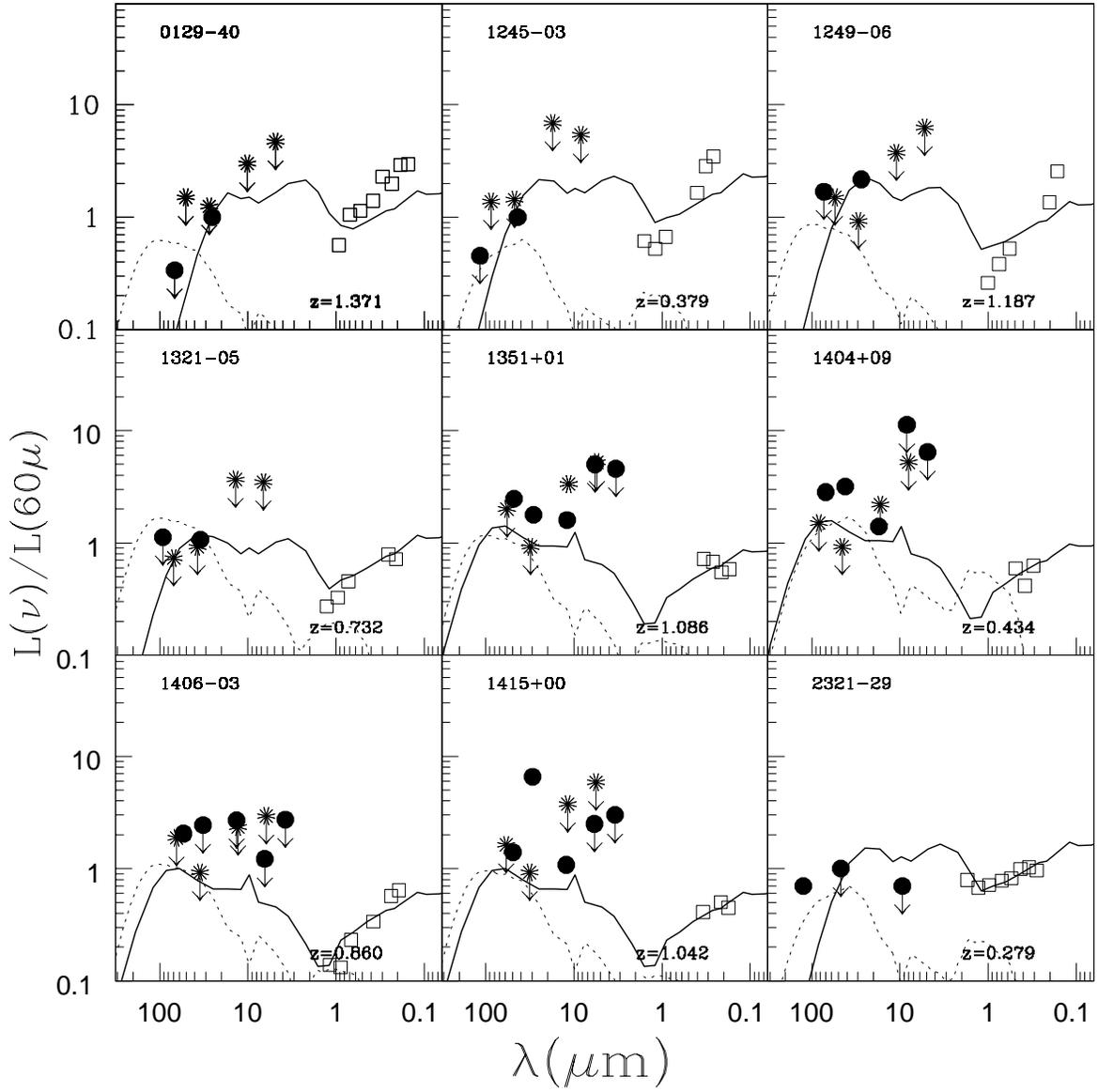}
   \caption{same as Figure \ref{fig:SED1}
              \label{fig:SED2}}
    \end{figure*}

   \begin{figure*}
   \plotone{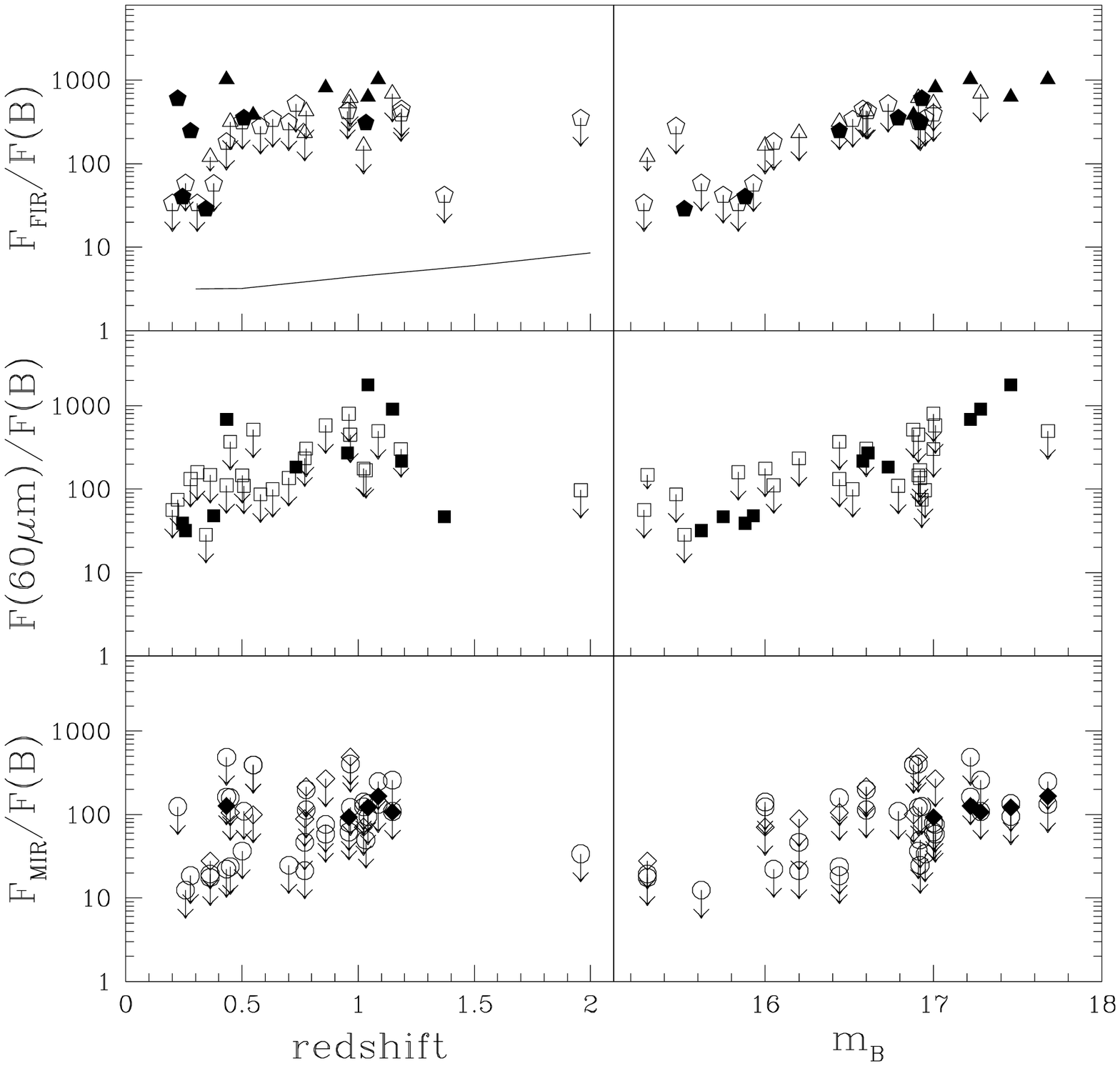}
	\caption{Logarithm of the ISOPHOT flux ratios versus
redshift (left panels) and B apparent magnitude (right panels).
The lower panels report the 7,12 and 25\,$\mu$m (MIR) fluxes, the upper
panels the 100 and 160\,$\mu$m (FIR) fluxes.
Symbols are as in Figure \ref{fig:fluxes}. A weak redshift-dependence is seen
in the left panels for long wavelength fluxes, while the ratio
does indeed show a dependence on m$_{\rm B}$. In the right panels
upper limits and detections are equally distributed.
The solid line in the upper left panel shows
the expected behaviour of the ratios
$\frac{\rm F_{\rm FIR}}{\rm F_{\rm B}}$ in presence of extinction.
%
\label{fig:colmB}
}
\end{figure*}

\section{Results}

\subsection{Notes on individual objects}

\begin{itemize}
\item
0051-39: has only one detection at 160 $\mu m$. The low value of the
60 $\mu m$ upper limit suggests that the 160 $\mu m$ flux could be eventually
affected by either a galactic cirrus or radio non-thermal emission.
The lack of radio data on this object does not allow us to check the latter
hypothesis.

\item
0059-30: has only one detection at 160 $\mu m$. This object is radio-loud
and therefore this measurement could be affected by a non-thermal component
but its flux is still compatible with the emission of a putative host
galaxy at this wavelength.

\item
0129-40: has a very clean 60\,$\mu$m detection, obtained with observations in
raster mode, confirming the higher sensitivity of this option.
\item
0144-39: detected both at 60 and 160 \,$\mu$m through raster pointings.
The 60\,$\mu$m IRAS flux agrees with the ISO detection, while that at
100\,$\mu$m is also in agreement with that at 160\,$\mu$m for a thermal
spectrum varying with wavelength as $\lambda^{3-4}$.

%
\item
1252+02: detected at 160 $\mu m$ and marginally at 60 $\mu m$.
IRAS found fluxes at 12 and 25 $\mu m$. It is very difficult to combine these
four measurements in a unique thermal component. They likely belong
to different thermal components.


\item
1355+02: detected at 60 $\mu m$ and possibily at 100 $\mu m$ by IRAS
with an upper limit at 160 $\mu m$. Its spectrum can be reproduced by the
emission of the circumnuclear torus.

\item
1404+09: has detections at 25, 60 and 100\,$\mu$m. Since it is a radio-loud
these measurements could be due to the combination of two different
emission mechanisms, thermal and non-thermal (see also \cite{pol00})
\item
1406-03, 1415+00 and 1415-00 have all large fluxes at 60 and 100
$\mu m$. The latter two were detected also at 25 $\mu m$.
 We cannot exclude that the large 60 $\mu m$ fluxes are
affected by instrumental artifact. Galactic cirrus can contribute
to the 100 $\mu m$ emission. The 100 $\mu m$ IRAS maps in these regions
show a quite high background of around 80 MJy/sr with a variation
from pixel to pixel around 10 \%. This means that 3$\pm$0.3 Jy fall
on the C100 pixel. An imperfect beam switching could then mimic a source
emission. However, we are not aware at present of such an effect in the
ISOPHOT data.
\item
2313-30: the only object observed both in raster and chopping modes but
detected only with the former instrumental set-up at 60 $\mu m$. The chopper
data are noiser but consistent with the raster detection.

\item
2321-29 and 2357-35 have a detection only at 160 $\mu m$. Without any other
FIR measurements it is tough to infer the nature of this flux. It could 
be ascribed to a cool component of the host galaxy as, for instance, seen
in \rg ~ (van Bemmel et al., 2000).

\end{itemize}

\subsection{Spectral Energy Distribution}

Figures \ref{fig:SED1} and
\ref{fig:SED2} show the spectral energy distribution
in the object's rest-frame of the 18 sources in the sample with FIR
detections. Spectra are normalized to the 60 $\mu m$ flux.
Open squares correspond to the UBVRI photometry and to JHK data,
filled circles to ISO measurements (down-arrows are
upper limits), stars to IRAS points, which are in general upper limits but for 
0105-26, 0120-28, 0144-39, 1252+02, 1351+01, 1355+02.

Although the spread in the SED appears large, mainly due to the presence
of censored data, some common features can be identified:
(a) all quasars with good near-IR data show a dip in the SED
around $\lambda = 1\,\mu m$
(b) all objects (but 0051-39) with 160 $\mu m$ data show 
a drop of the far-IR SED beyond 100 $\mu m$.

The comparison between the optical and far-IR spectrum
must take into account
possible variations due to the intrinsic variability of these sources.
The effect of the variability is difficult to address in detail, as it
significantly depends on the time lag between the various flux measurements, the
quasar absolute luminosity and increases with decreasing wavelength. From the
analysis of the structure function \citep{Cri96} we should
expect an average uncertainty due to variability of 0.2 magnitudes for quasars
having a typical absolute magnitude M$_B\sim$-26 (0.1 for
M$_B\sim$-30 and 0.3 for M$_B\sim$-23). Such variations of the
flux values are still within the size of the photometric points shown
in Figures \ref{fig:SED1} and \ref{fig:SED2}.

The thick solid curves in Figures \ref{fig:SED1} and \ref{fig:SED2}
represent the predictions of the Granato and Danese 's model (1994), which
computes the expected emission of a circumnuclear torus-like dust
distribution around the central nuclear source. The solution
of the radiative transfer equation is required since in such tori the dust
emission is self--absorbed even in the near-- and mid--IR.
\hfill\break
Dust is illuminated by the central source and reaches an equilibrium 
temperature which is a function of the intensity of the radiation field
and of the dimensions and chemical composition of the assumed grain components.
In the case shown here the chosen geometry for the dust distribution
has an axial symmetry and
a scaleheight along the vertical axis increasing linearly with distance
(`flared disc').
The central source is either unobscured or only partially obscured from view.
This model naturally explains the SED dip around 1\,$\mu$m: dust too close
to the central engine attains too high temperature and sublimates.

All the curves shown in Figures \ref{fig:SED1} and \ref{fig:SED2} correspond
to the same 'face-on' model, being all these UV-excess objects with
the central source un-extincted by the surrounding torus material.
The difference in the amount of IR fluxes can be accounted for by varying
the dust distribution size, i.e. the dimension of the torus.\hfill\break
\hfill\break
We have neither attempted to find the best-fit curves,
because of the lack of sufficient constraints from the data, nor
we claim that the model shown is the 'best' to explain these observations.
Other models would be equally good (see e.g. Rowan-Robinson, 1995).
\hfill\break
However, the comparison between the predicted curves and the photometric
points provides useful insights:
a simple torus-like dust distribution around the central source is 
a too naive picture. In some cases the required dimension
of the dust distribution is wide and dust could be broadly spread
in a putative host galaxy disc.
\hfill\break
To show how an additional galaxy disc emission modifies the SED,
the optical-{\it mm} spectrum of M82, a
typical local starburst, is also plotted in Figures \ref{fig:SED1}\
and \ref{fig:SED2}, as a dotted line.

   \begin{figure}
   \plotone{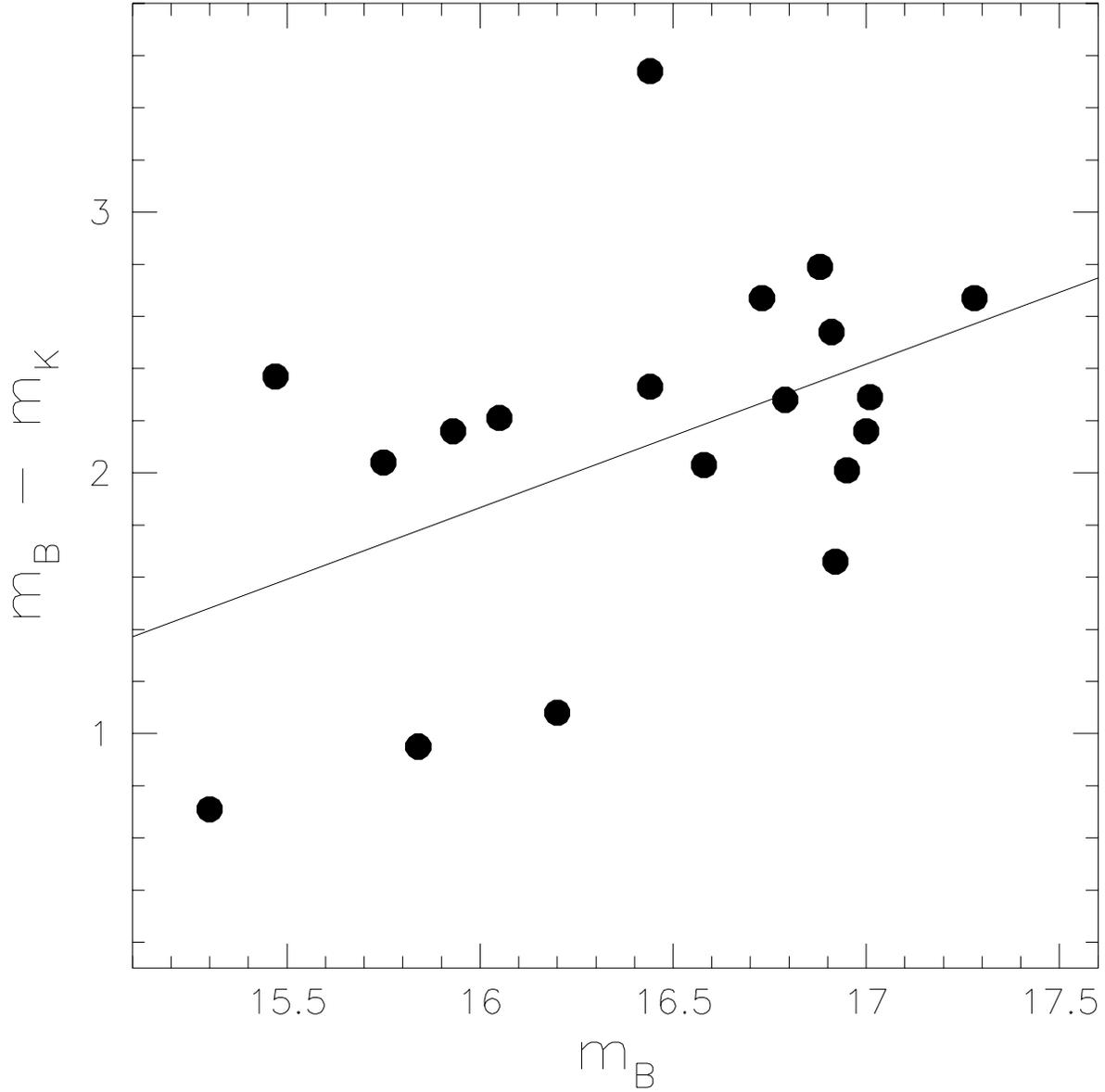}
\caption{Flux ratios between the K-band and the B-band,
expressed as ${\rm m_B} - {\rm m_K}$, against the
 B apparent magnitude for those objects with near-IR photometry reported
in Table 1. {\bf The solid line corresponds to the estimated
linear regression line, the standard deviation, 0.575, is large and the correlation
coefficient is 0.33. The hypothesis that $\frac{\rm F_K}{\rm F_B}$ does not
vary with m$_{\rm B}$ can only be rejected at a 13\% level.}
\label{fig:mBmK}
}
\end{figure}

   \begin{figure}
   \plotone{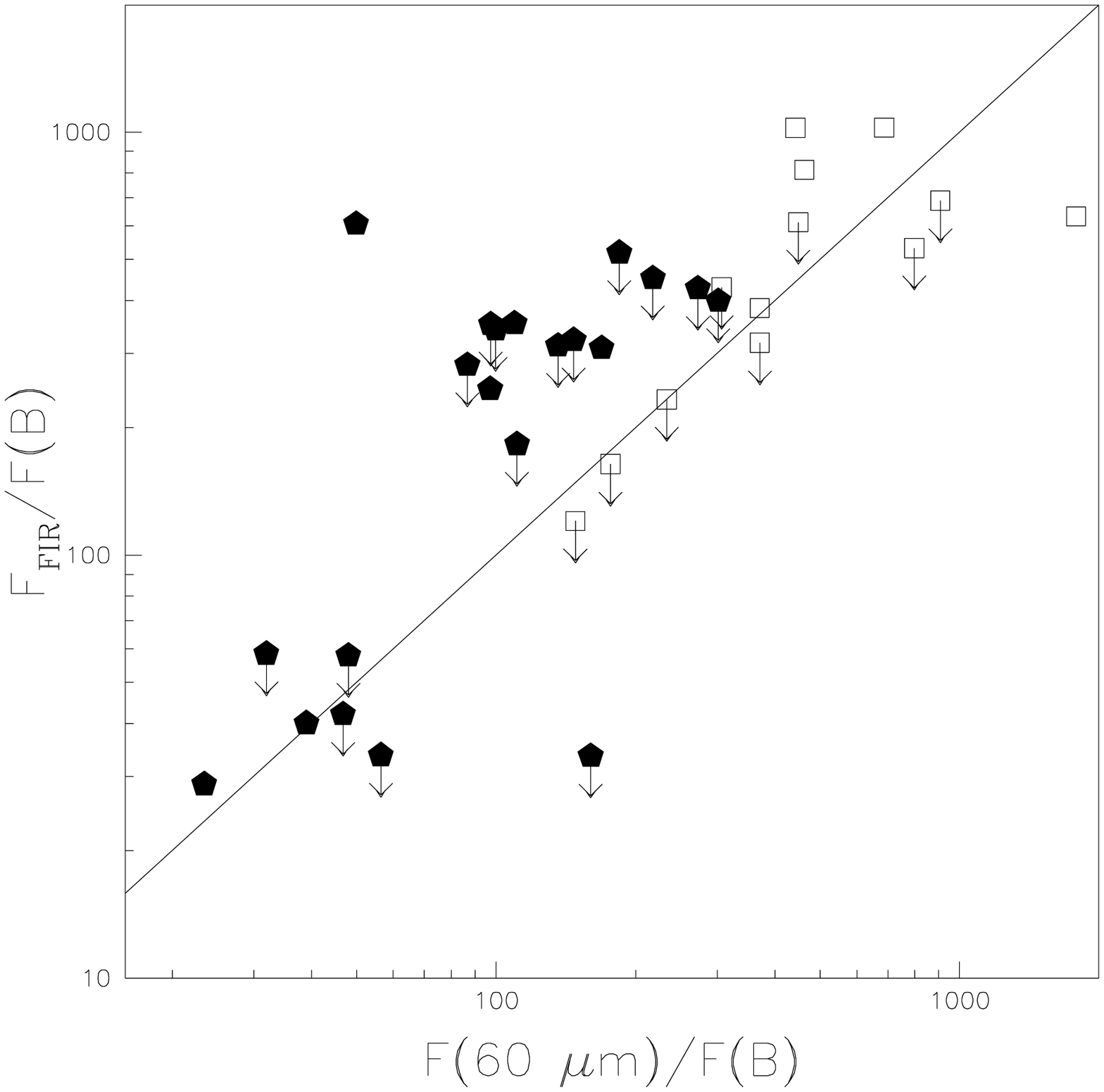}
\caption{The ratio between the 100 and 160 $\mu m$ fluxes and
the B-band
flux is shown as a function of the ratio between the 60 $\mu m$ and the
B flux. Open squares correspond to the 100 $\mu m$ data while filled pentagons
to the 160 $\mu m$ ones. The straight line corresponds to a one-to-one
relationship between the two colours. The outlier at
$\frac{\rm F_{160}}{\rm F_B} \sim 600$ and
$\frac{\rm F_{60}}{\rm F_B} \sim 50$ corresponds to 0051-39, which, as shown
in Figure \ref{fig:SED1}, has an extraordinarly large 160$\mu m$ flux.
\label{fig:col3}
}
\end{figure}

\subsection{The colour plots} 

Figure \ref{fig:colmB}
reports the colours, $\frac{\rm F_{\rm MIR}}{\rm F_{\rm B}}$,
$\frac{\rm F_{60\mu m}}{\rm F_{\rm B}}$,  $\frac{\rm F_{\rm FIR}}{\rm F_{\rm B}}$
as a function of the B-band apparent magnitude (right panels) and of the
source redshift (left panels).\hfill\break
The increase of the ratios with the B-magnitude 
can be due to several effects:
(a) an observational bias, (b) cirrus emission or (c) extinction.

(a) The upper left corner of every figure panel would contain objects with
large far-IR emission and bright apparent blue magnitude and therefore
easily observable.
Objects in the lower right corner could be missed
because of a lack of sensitivity of far-IR observations.
The upper limits distribution is quite uniform with respect to m$_B$,
i.e. there is no accumulation of upper limits at the faint end.
Sensitivity limits therefore do not seem to affect dramatically the results.
\hfill\break
If the variation of the ratios in Figure~\ref{fig:colmB}
were only linked to the QSO spectrum, we
would expect to see this ratio decreasing with redshift:
the B-flux increases with redshift because of the negative k-correction and
for a constant FIR flux $\frac{\rm F_{\rm FIR}}{\rm F_{\rm B}}$ would be
smaller at higher redshifts \footnote{k-corrections of FIR fluxes are
neglected here since, within the redshift range of our objects,
the observed FIR fluxes sample the Wien region of the thermal spectrum
and the resulting variation of the FIR fluxes in the restframe and observed
frame is small.}. The observations show the opposite trend.

(b) The presence of Galactic cirrus emission falling in the ISOPHOT beam
and affecting the far-IR fluxes has been checked using independent measurements
at these wavelengths. We used the IRAS 100 \,$\mu$m maps in regions
surrounding the observed objects (see Appendix for detail).
%
Although in three cases the far-IR fluxes can be really enhanced because of a
cirrus contribution (see Appendix)
there is not a one-to-one correspondence between the ratios
$\frac{\rm F_{\rm FIR}}{\rm F_{\rm B}}$ and the 100\,$\mu$m rms values.
 
(c) The observed behaviour could be then ascribed to
an extinction along the line of sight.
The left upper panel shows a small increase
of $\frac{\rm F_{\rm FIR}}{\rm F_{\rm B}}$ with redshift. The solid line
corresponds to the expected change of this ratio when the
emitted UV-optical flux (${\rm F}(\lambda) \propto \lambda ^{-0.7}$)
varies, because of the extinction $A(\lambda)$ 
and k-correction, $k(\rm z)$, as
\begin{equation}
{\rm F_B}
= {\rm F}(\lambda, {\rm z}) k({\rm z}) \cdot 10^{-0.4 \cdot A(\lambda)}
\label{eq:ext}
\end{equation}
\hfill\break
We compute the ratio $\frac{\rm F_{\rm FIR}}{\rm F_{\rm B}}$
for blue fluxes, ${\rm F _B}$, affected by an extinction of
$A(\lambda) = E_s(B-V)* Q_{ext}(\lambda)$. We adopted as extinction
law that provided by Calzetti (2001) for starburst galaxies.
\hfill\break
Neglecting k-correction of the far-IR fluxes and
correcting B-fluxes as in eq.\ref{eq:ext}, more than two third
of the ratios, $\frac{\rm F_{\rm FIR}}{\rm F_{\rm B}}$, in the upper right
panel of Fig.\ref{fig:colmB} would be shifted to an average value of 100.
This confirms that the gross behaviour of $\frac{\rm F_{\rm FIR}}{\rm F_{\rm B}}$
is dictated by extinction.


Our interpretation is therefore that {\it fainter objects in the optical are
also redder}.
The redshift-dependence of IR to blue flux ratios were also investigated
by Alexander et al. (2001) for a sample of ELAIS AGN with 15\,$\mu$m fluxes.
These authors find a ratio $\frac{f_{\rm MIR}}{f_{\rm B}}$ ranging between
2.5 and 100 independent of the source redshift.
Haas et al. (2000) as well do not detect any variation of both
$\frac{L_{\rm MIR}}{L_{\rm B}}$ and $\frac{L_{\rm FIR}}{L_{\rm B}}$
with redshift.
\hfill\break
To further check the presence of dust extinction along the line of sight
through the sources, we plot in Figure \ref{fig:mBmK} the ratio between
the K-band and the B-band fluxes, expressed as ${\rm m_B} - {\rm m_K}$
against the B magnitude. {\bf The straight line in Figure \ref{fig:mBmK}
represents to the estimated linear regression line. In spite of the large
spread (standard deviation is 0.575), there is a marginal correlation
(the correlation coefficient is 0.33) and a slight probability
that at larger values of ${\rm m_B}-{\rm m_K}$ correspond larger m$_B$
in accordance with the ratios plotted in
Figure \ref{fig:colmB}. However, the hypothesis that
$\frac{\rm F_K}{\rm F_B}$ does not
vary with m$_{\rm B}$ can only be rejected at a 13\% level.}

\subsubsection{Colour-colour plot} 

In Figure \ref{fig:col3}, 100 and 160 $\mu m$ fluxes,
normalized to the B-band flux, $\frac{\rm F_{\rm FIR}}{\rm F_{\rm B}}$,
are shown against the 60 $\mu m$ flux, $\frac{\rm F_{\rm 60}}{\rm F_{\rm B}}$.
The two colours are tightly
related: an increase of the long wavelength flux with respect to the
B-band flux corresponds to an equivalent enhancement of the 60
$\mu m$ flux with respect to the optical one. The two far-IR fluxes
are therefore related and very likely belong to the same thermal
component. As a comparison a straight line showing a one-to-one relationship
is drawn in Figure \ref{fig:col3}.
The ratio $\frac{F(60\mu m)}{F_{\rm FIR}}$ can be then used
as an estimation of the dust colour temperature for that thermal component.

\subsection{Dust Temperature}

\begin{figure} 
   \plotone{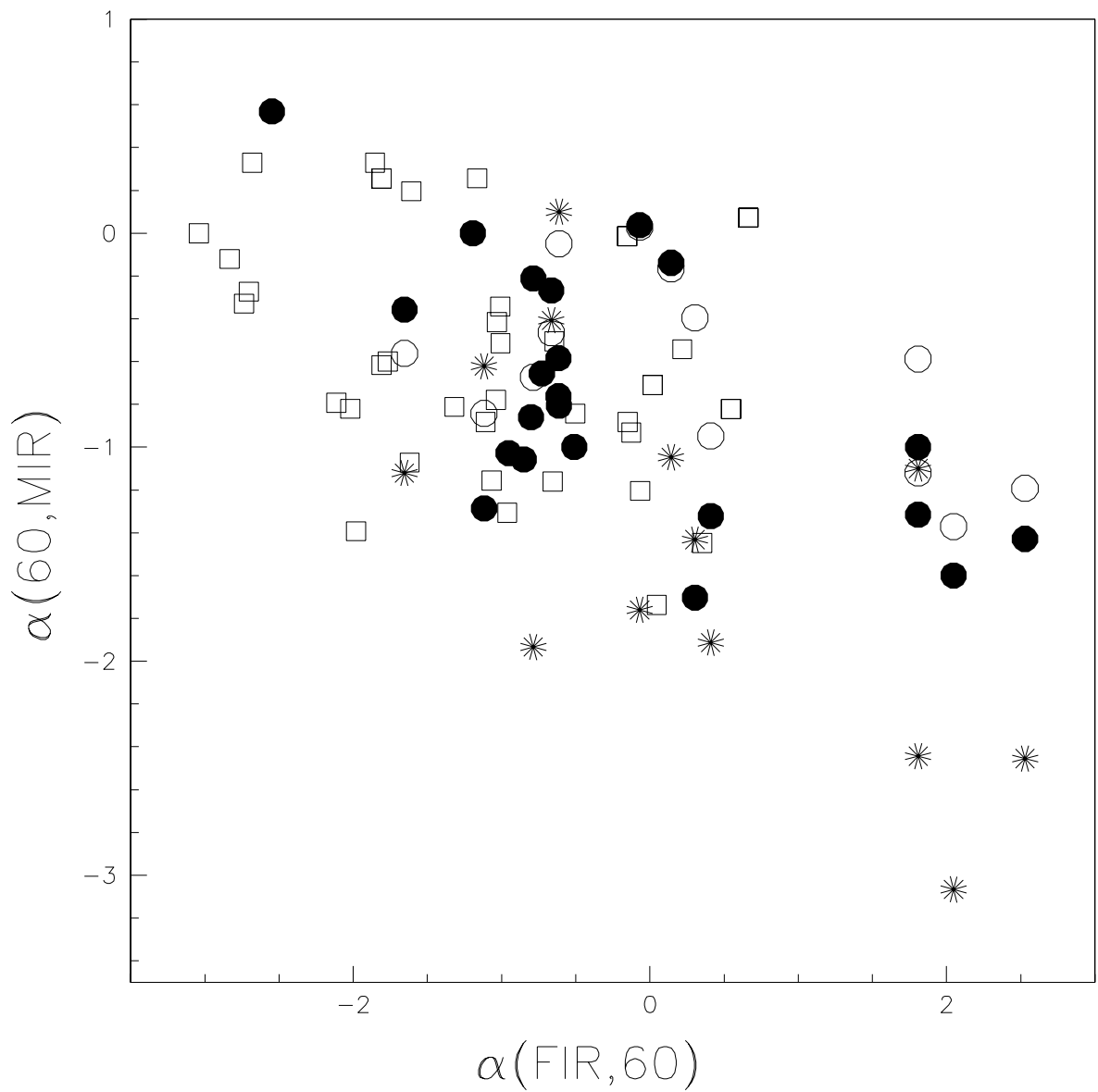}
   \caption{The ratio between 60$\mu$m flux
and MIR fluxes
is shown as a function of the ratio between FIR and 60 $\mu m$ fluxes.
Ratios are defined by equation \ref{eq:index}.
Asterisks correspond to the 25 $\mu$m data, filled circles
to 12\,$\mu$m, open circles to 7\,$\mu$m ones. Open squares correspond
to QSOs at $z \leq 2$ reported in Andreani, Franceschini \& Granato (1999).
Distinct areas can be
identified from AGN dominated (in the upper left corner ) to
Starburst dominated objects (in the lower right corner), according to
the expected dominant emitting mechanism in the object SED from
non-thermal to thermal.
\label{fig:col2}
}
\end{figure}

Customarily the spectral indices, $\alpha$, for a $f_\nu \propto \nu^\alpha$
SED are defined as
\begin{equation}
\alpha(\nu_1,\nu_2)=\log \frac{f(\nu_1)}{f(\nu_2)} -\log\frac{\nu_1}{\nu_2}
\end{equation}
\label{eq:index}
\noindent
and simply related to colours. Figure
\ref{fig:col2} reports $\alpha(60\mu m,{\rm MIR})$ versus
$\alpha({\rm FIR},60\mu m)$. According to our previous finding (\S 4.3)
this latter is inversely proportional to the {\it warm}
dust temperature, T$_{\rm wd}$.
Although many points in the diagram are upper/lower limits
there seems to be a trend: the relative emission at shorter wavelengths
(7, 12 and 25 $\mu m$) with respect to the 60$\mu $ emission
decreases as T$_{\rm wd}$ increases.
This colour-colour diagram was widely used in the past
as a tool to detect and discriminate between different types of activity
in the nuclear and circumnuclear regions of galaxies
(see e.g., Canalizo \& Stockton (2001) and references therein).
Different kinds of objects (QSOs/Seyferts, starbursts and ULIRGS) occupy
distinct loci in such a diagram: objects for which FIR fluxes are dominated
by dust
reradiation are located in the lower right corner (in our
convention at lower T$_{\rm wd}$ and lower $\alpha(60\mu m,{\rm MIR})$),
while objects with strong non-thermal emission, such as optically selected QSOs,
are in the upper left corner. Indeed PG quasars observed by IRAS are
preferentially found at $-2 < \alpha(60\mu m,{\rm MIR}) < 1$ and
$-2 < \alpha({\rm FIR},60\mu m) < 0$ partially overlapping the region of the
diagram where {\it transition} objects are found
(Canalizo \& Stockton, 2001).
To show that in detail we plot in Figure \ref{fig:col2} the location of the
QSOs observed by IRAS at redshift lower than 2 and
studied by Andreani, Franceschini \& Granato et al. (1999). 

\begin{figure}
   \plotone{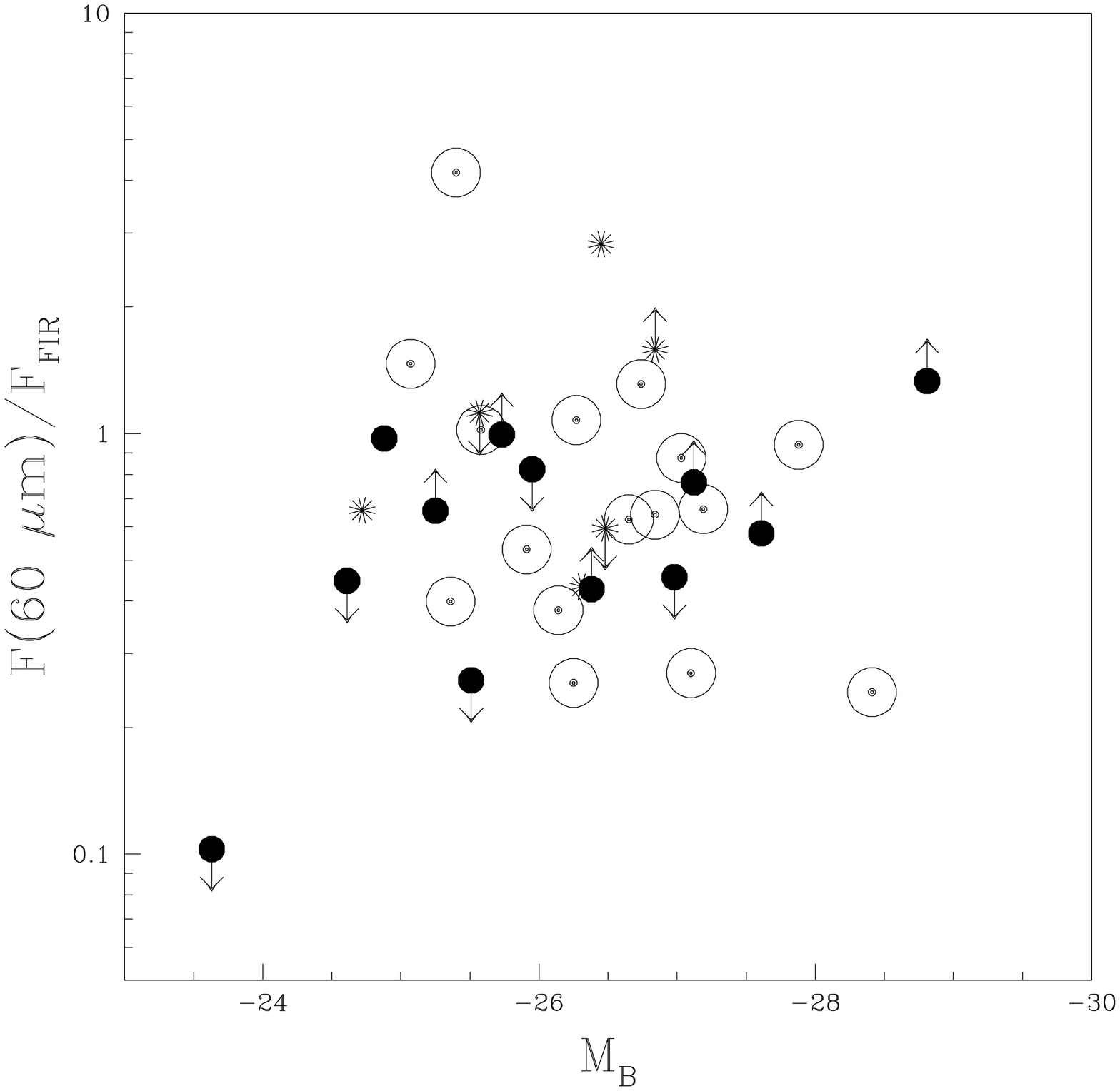}
	\caption{The ratio $\frac{F(60\mu m)}{F_{\rm FIR}}$, which is
is proportional to the dust temperature, T$_{\rm wd}$, as a function of
the absolute B magnitude, M$_{\rm B}$. Asterisks refer to the 100\,$\mu$m data,
filled circles to the 160\,$\mu$m data, $\odot$ symbols to ratios
with upper limits on both fluxes.
\label{fig:assmag}
}
\end{figure}

In Figure \ref{fig:assmag} the ratio $\frac{F(60\mu m)}{F_{\rm FIR}}$
is plotted against the absolute B magnitude. 
With all the limits of this analysis due to the
poor information on the FIR fluxes, it appears that no statistically significant
correlation is present between the B luminosity and the warm dust colour
temperature, T$_{\rm wd}$.
\hfill\break
If M$_{\rm B}$ is linked to the nuclear emission, while
$\frac{F(60\mu m)}{F_{\rm FIR}}$ is only a dust property, it seems that no
strong relation exists between the energy emitted by the nuclear source and that
emitted in the FIR. Similar findings were already reported by
Andreani, Franceschini \& Granato (1999)
for an inhomogeneous sample of optical
quasars with IRAS and mm-wavelengths fluxes. For their sample
no relation was found between the dust mass and the absolute 
B-magnitude M$_{\rm B}$. 
McMahon et al. (1999) and Omont et al. (2001)
do not detect any dependence of the sub-mm fluxes on the optical luminosity
in high redshift quasars. All these results could be fitted into a picture
in which the emission from the nuclear source is not
tightly related to the physics (the dust properties: temperature, mass and
luminosity) of the surrounding medium and argues in favour of
the hypothesis that most
of the FIR luminosity is linked to a concurrent starburst and trace
activity in the host galaxy
(see also Clements, 2000). This result is also related to dynamical
studies aiming at the search
for BHs in galaxy centres. These investigations show that although there is
a strong one-to-one correspondence between the central BH mass and bulge
luminosity no detectable correlation is found between the BH mass
and the luminosity of galaxy discs (see e.g. Kormendy \& Gebhardt,2001).


   \begin{figure*}
   \plotone{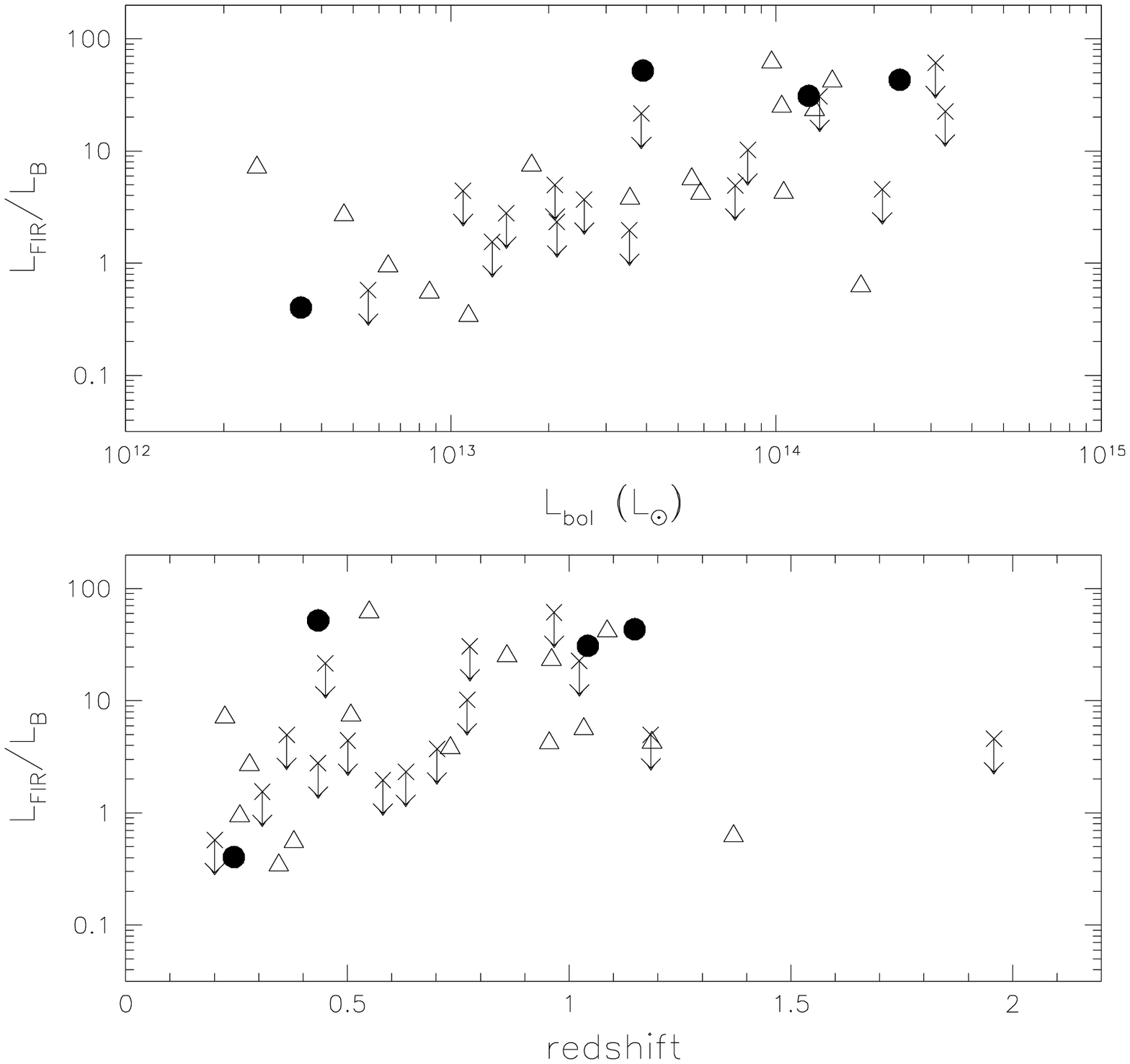}
	\caption{
The ratio between the far-IR, $\rm L_{\rm FIR}$, 
and the B-band, $\rm L_{\rm B}$,
luminosities against the bolometric luminosity, $\rm L_{\rm bol}$ (upper panel)
and against the source redshift (lower panel) is shown.
Triangles correspond to those objects with only one ISOPHOT detection,
filled circles to those with two or three detections, downarrows to
undetected objects.
\label{fig:lum}
}
\end{figure*}

\subsection{Luminosity plots}

Figure \ref{fig:lum} shows the ratio between the far-IR, $\rm L_{\rm FIR}$, 
and the B-band, $\rm L_{\rm B}$,
luminosities versus the bolometric luminosity, $\rm L_{\rm bol}$. 
On the y-axis we plot $\frac{\sum _i \nu_i f(\nu _i)}{\nu_B f(\nu_B)}$ 
i.e. the ratio between the far-IR luminosity
$\rm L_{\rm FIR}$, as the sum of the different far-IR data,
and the blue-band luminosity computed from the absolute
B magnitude: $10^{-0.4\cdot \rm M_{\rm B} + 28.523}$. On the x-axis
the bolometric luminosity is shown:

\begin{equation}
L_{\rm bol} = 4\pi d^2_L ~ \int d\nu ~f(\nu)
\end{equation}

\noindent
where $d_L$ is the luminosity distance. L$_{\rm bol}$ is computed
from UV to FIR.
Down-arrows correspond to objects undetected in all bands, triangles to
those with one detection, filled circles to those detected at least in
two bands. $L_{\rm FIR}$ could well be
underestimated for those objects with only one or two detected fluxes.
Sampling the Wien region of a thermal spectrum with a characteristic temperature
of 40-100 K at only one or two
frequencies means to underestimate the total luminosity of a factor between
2 and 5 with respect to the integral over the whole frequency range.

Only five objects have clearly a $\frac{\rm L_{\rm FIR}}{\rm L_{\rm B}}$
ratio lower than 1, 14 detected ones have
$\frac{\rm L_{\rm FIR}}{\rm L_{\rm B}} > 1$ of which 7
have FIR luminosity more than one order of magnitude higher than the blue one.
There is a clear trend of increasing
$\frac{\rm L_{\rm FIR}}{\rm L_{\rm B}} $
with increasing bolometric luminosity. This means that the FIR luminosity
becomes increasingly important as the bolometric luminosity increases.
A similar behaviour is seen in nearby ultraluminous IR Galaxies (ULIRGs)
(Sanders and Mirabel, 1996) and has been associated to the increased central
concentration of molecular gas. Because of the high extinction
the relative role of nuclear starburst and AGN activity is hard
to disentangle in ULIRGs but there seems to be
an enhancement of AGN contribution to the overall energy budget
as the bolometric luminosity increases.
Furthermore, in the Unified Scheme of AGN an enhancement
of the $\frac{\rm L_{\rm FIR}}{\rm L_{\rm B}} $ ratio
is expected, since the reprocessed emission by the
dusty torus emerges isotropically in the FIR.
The colour axis $\frac{\rm L_{\rm FIR}}{\rm L_{\rm B}} $ should be
related to the viewing angle to the torus with more inclined objects
having larger $\frac{\rm L_{\rm FIR}}{\rm L_{\rm B}} $
therefore lying on the right-hand side of the diagram.

{\bf The lower panel in Figure \ref{fig:lum} shows the behaviour of the
ratio $\frac{\rm L_{\rm FIR}}{\rm L_{\rm B}} $
against the source redshift: with the present data it is not possible to detect any evolution of the ratio $\frac{\rm L_{\rm FIR}}{\rm L_{\rm B}} $ with
redshift.}

\section{The quasar composite spectrum}
The objects presented in this work are randomly drawn from a complete
sample of optically selected quasars (see \S 2) and can be
considered representative of the entire quasar population.
If we assume that the radio-quiet quasars can be modeled as
a homogeneous population, a spectrum can be built
via observations at fixed frequencies of targets at different redshifts.
This analysis exploits the whole spectral information from the optical to
the FIR and helps outlining possible general features of the emitting
mechanisms and the physics of the QSO environment.
The spectrum is built by dividing the available wavelength range in bins.
In each bin an average of the fluxes at that wavelength is computed.
The errorbars shown are those related to the averages. Uncertainties on the
flux scale lie well within the large errorbars.

\begin{figure}
   \plotone{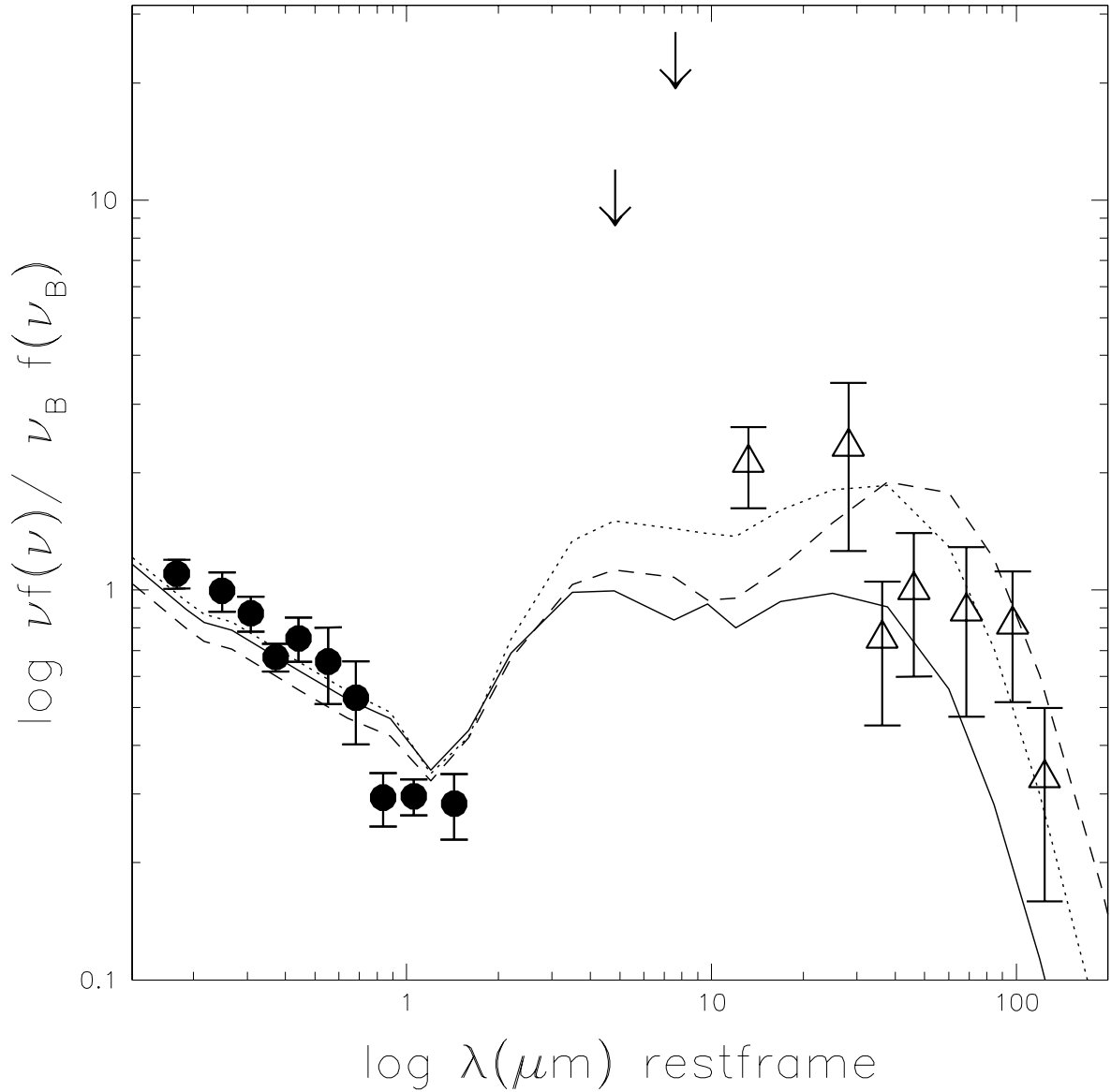}
	\caption{
The average spectrum in the QSO restframe.
The spectrum is normalized at $\lambda _{\rm B} =0.44 \,\mu$m.
Filled circles correspond to the optical and near-IR data, open triangles to
the ISOPHOT measurements discussed in this paper. The lines going through the
data show the predictions of the models by Granato and Danese (1994)
which are able to fit the QSO SED (see \S 4.2). 
The solid line corresponds to a face-on compact configuration of the torus
and absorption at 30$\mu$m of $\tau$=30, dotted line to a 45$^\circ$
inclined torus with larger dimension and same $\tau$, the dashed line to a
45$^\circ$ inclined torus with higher absorption, $\tau$=60.
\label{fig:compsp}
}
\end{figure}

Figure \ref{fig:compsp} shows the average spectrum in the QSO restframe.
The spectrum is normalized at $\lambda _{\rm B} =0.44 \,\mu$m.
Averages are computed in each wavelength bin with the survival
statistics which take into account the censored data \cite{fei}.
Detections with a suspected contribution from IR-cirrus were included as upper limits.
A well defined IR component peaking at 10-30 $\mu m$ and dropping steeply
above 100\,$\mu$m is evident.
The spectral shape of the far-IR component is characteristic of a thermal
origin. It can be speculated whether it is due to starburst emission in
the host galaxy and/or to a dusty torus around the
central source.
In order to derive constraints to the available parameters on the dust
distribution, we can compare the data with available models.
Figure
\ref{fig:compsp} reports three different spectral energy
distributions predicted by the Granato \& Danese 's model (1994) (see \S 4.2).
We restrict ourselves to models which allow UV-optical photons to
escape from the nuclear
source since the quasars were selected on the basis of their UV-excess.
The dust distribution around the central source should be compact
because of the sharp cut-off at long wavelengths of the FIR bump.
A larger bump would imply larger temperature distribution
and very likely larger spatial dimension of the dust distribution.

The solid line corresponds to face-on (with a 'naked' nuclear source)
compact (i.e. the ratio between the inner and outer radius of the
torus $\frac{r_{in}}{r_{out}}$=800)
configuration of the torus and an optical depth $\tau$ at 0.3 $\mu m$
of 30; the dotted line to a 45$^\circ$ inclined torus with 
larger dimension ($\frac{r_{in}}{r_{out}}$=1000)
and same $\tau$, the dashed line to a 45$^\circ$ inclined
torus ($\frac{r_{in}}{r_{out}}$=1500) with higher absorption, $\tau$=60.
The curve in better agreement with the available photometric points
is that corresponding to the larger optical depth (dashed line)
and therefore the strongest dust self-absorption around 10-30\,$\mu$m.

\section{Conclusions}

A sample of optically selected quasars was observed with ISOPHOT
between 7 and 160 $\mu m$, with the following results:

\begin{itemize}

\item
The sample shows homogeneity in
mid and far-IR properties: the mid and far-IR colours depend neither on
the redshift nor on the absolute B-magnitude and there is no correlation
between the blue magnitude and the ISOPHOT fluxes.

\item
The detection rate is not a function of redshift (
10 detected objects have $z<0.7$, the other 10 have $z>0.7$, where 0.7 is the
median redshift of the redshift distribution)
consistently with the negative
K-correction implied by a thermal spectrum.

\item
The long wavelengths ($\lambda$=60,100 and 160\,$\mu$m)
fluxes are tightly related and likely arise from the same thermal
component.

\item
Objects with fainter blue magnitudes, m$_{\rm B}$, have larger ratios
between the FIR ($\lambda > 60 \mu m$) fluxes and the blue band flux,
$\frac{f_{\rm FIR}}{f_{\rm B}}$. We ascribe this behaviour
to a larger extinction along the path to the central source.
\item
The colour-colour diagram
$\alpha(60\mu m,{\rm MIR})$ versus $\alpha({\rm FIR},60\mu m)$
shows that the objects of the
present sample span a wide range of properties from AGN-dominated
ones (low $\alpha({\rm FIR},60\mu m)$, i.e.
large dust temperature) to starburst-dominated ones
(low $\frac{60\mu m}{\rm MIR}$ and large  $\frac{\rm FIR}{60\mu m}$).

\item
No statistically significant correlation is detected between the
absolute B-magnitude, M$_{\rm B}$ and the warm dust temperature,
T$_{\rm wd}$.
If M$_{\rm B}$ mainly reflects the nuclear emission and the FIR fluxes
the emission of the surrounding dust, it seems that no relation exists between the
energy emitted by the nuclear source and that by dust. The latter could
rather 
be a tracer of the starburst activity in the host galaxy.
\item  
Even in an optically selected sample the amount of energy emitted in the far-IR,
$\rm L_{\rm FIR}$, is on average a few times larger than that
emitted in the blue, $\rm L_{\rm B}$. The ratio $\frac{L_{\rm
FIR}}{L_{\rm B}}$ does not show any clear variation with redshift but
increases with the bolometric
luminosity, L$_{\rm bolo}$.
This behaviour is similar to that seen in local ULIRGs, possibly
because of
an increased central concentration of molecular gas.
In the Unified Scheme of AGN an enhancement
of the $\frac{\rm L_{\rm FIR}}{\rm L_{\rm B}} $ ratio
is expected in the more inclined objects for which the
viewing angle goes through the torus, i.e. through a more extincted line of
sight.
\item
The QSO Composite Spectrum, i.e. the average spectrum built in the QSO
restframe, shows a broad far-IR bump at 10-30 $\mu m$ which
could be due either 
to a starburst emission in the host galaxy or to a dusty torus around the
central source as predicted in the unified scheme of AGN.
\hfill\break
Restricting our analysis to only the QSO SED one could argue that
the best picture fitting the data is that of a circumnuclear dusty torus.
However other arguments presented also in this paper (the colour-colour
diagram, the lack of relation between the power of the nuclear source and
the dust temperature and the increase in the ratio
$\frac{\rm L_{\rm FIR}}{\rm L_{\rm B}} $, see
Figures \ref{fig:col2}, \ref{fig:assmag}, \ref{fig:lum}) suggest that
a contribution from
a starburst from the host galaxy is quite unavoidable.
\end{itemize}


\acknowledgments

This publication makes use of data products from the Two Micron All Sky Survey, 
which is a joint project of the University of Massachusetts and the Infrared
Processing and Analysis Center/California
Institute of Technology, funded by the National Aeronautics and Space
Administration and the National Science Foundation.
\hfill\break
Based on photographic data obtained using The UK Schmidt Telescope. The
UK Schmidt Telescope was operated by the Royal Observatory Edinburgh,
with funding from the UK Science and
Engineering Research Council, until 1988 June, and thereafter by
the Anglo-Australian Observatory. Original plate material is
copyright the Royal Observatory Edinburgh and the
Anglo-Australian Observatory. The plates were processed
into the present compressed digital form with their
permission. The Digitized Sky Survey was produced at the
Space Telescope Science
Institute under US Government grant NAG W-2166.
\hfill\break
One of the authors (PA) warmly thanks Ilse van Bemmel for her help during the
data analysis and for her permission of using her software. 
PA acknowledges the fundamental
help of the ISOPHOT Data Center in Heidelberg in the data reduction
and MPE for hospitality.
Alexander von Humboldt Foundation for support is acknowledged.
Part of this work was funded by the Italian Space Agency
(ASI) under contract ARS-98-226, ASI-I-R-105-2000.

\appendix

\section{Appendix: Detections against cirrus level}

\begin{figure}
   \plotone{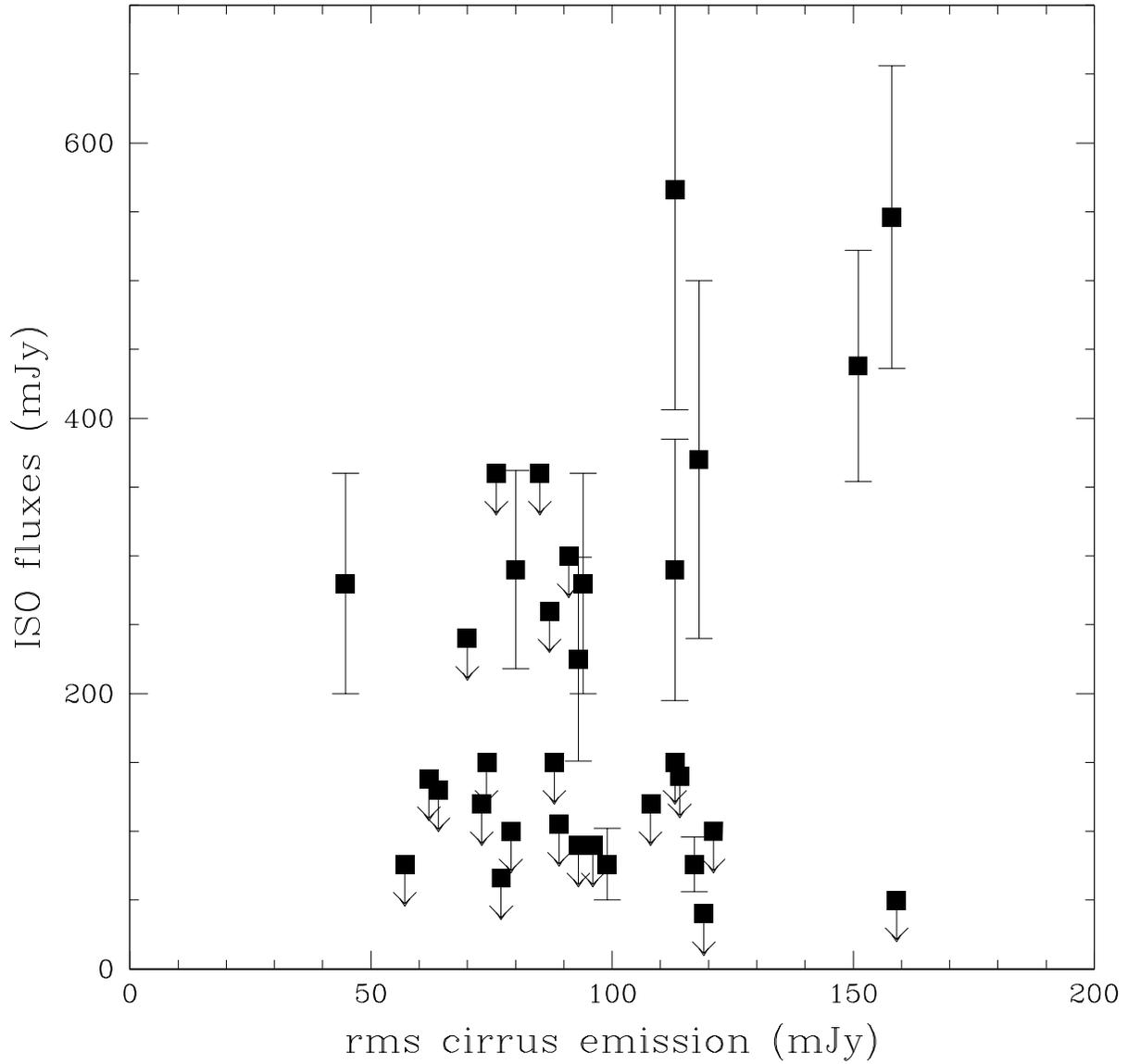}
	\caption{
The 100 and 160\,$\mu$m fluxes against the rms values of the Galactic cirrus
emission at 100\,$\mu$m as estimated from the IRAS maps towards the sources
observed by ISO. Upper limits are indicated as downarrows. The three higher
values of the ISO fluxes correspond to 0051-30, 1404+09 and 1406-03, the first
two are also radio-loud objects.
\label{fig:appendix}
}
\end{figure}

Figure \ref{fig:appendix}
reports the detected fluxes and upper limits at
100 and 160\,$\mu$m against the r.m.s. values of the cirrus emission
as estimated from the 100\,$\mu$m IRAS maps in the sky positions
corresponding to the source location. Upper limits are shown by downarrows.
No correlation between the two quantities is apparent,
with the possible exception of the three sources 0051-39, 1404+09, 1406-03.
For these three cases the large observed flux could be affected by cirrus
emission as discussed in section 4.1. However, 0051-39, 1404+09 are
radio-loud objects and the tail of the radio emission
could also contribute to the FIR fluxes.

\clearpage


\makeatletter
\def\jnl@aj{AJ}
\pagestyle{empty}
\ifx\revtex@jnl\jnl@aj\let\tablebreak=&\nod\nl\fi
\makeatother
\def\nod{\nodata}
\voffset=-1in
\hoffset=-1in
\begin{deluxetable}{cccrrrrrrrrrrr}
\tablewidth{48pc}
\tablecaption{Optical and near-IR Photometry of Quasars}
\tablehead{
\colhead{Name}     &\colhead{R.A.(J2000)} &
\colhead{dec(J2000)} &
\colhead{z}      &
\colhead{m$_{\rm B}$}          & \colhead{m$_{\rm U}$}  &
\colhead{m$_{\rm V}$}          & \colhead{m$_{\rm R}$}  &
\colhead{m$_{\rm I}$}          & \colhead{m$_{\rm J}$}  &
\colhead{m$_{\rm H}$}          & \colhead{m$_{\rm K}$}  &
\colhead{M$_{\rm B}$}          & \colhead{P$^{~a}_{\rm rad}$}
}
\startdata
 0034-70 & 0 34 05.0 &-70 25 54 &0.363 &15.30 &\nod & 15.50& 17.70&\nod  &
16.10&15.61&14.59&-26.27&25.8\nl
 0049-29 & 0 51 38.1 &-29 23 13 &0.308 &15.84 &16.66& 14.81& 13.96& 13.59&
 14.99&14.37&14.89&-25.40&\nod\nl
 0051-39 & 0 54 09.5 &-39 16 52 &0.224 &16.93 &16.92& 17.30& 16.62& 16.51&
\nod  &\nod &\nod &-23.63&\nod\nl
 0059-30 & 1 02 00.7 &-30 18 28 &1.033 &16.92 &16.41& 17.01& 16.64&\nod  &
16.11&15.77&15.26&-26.98&26.7\nl
 0105-26 & 1 05 58.9 &-26 06 22 &0.776 &16.60 &\nod & 16.90& 15.60&\nod  &
\nod  &\nod &\nod &-26.65&\nod\nl
 0120-28 & 1 22 36.6 &-28 43 20 &0.434 &16.05 &15.48& 15.71& 15.84& 15.91&
15.48&14.68&13.84&-25.91&\nod\nl
 0129-40 & 1 31 43.9 &-40 36 54 &1.371 &15.75 &15.06& 15.72& 15.13& 15.16&
 14.68&13.89&13.71&-28.81&\nod\nl
 0144-39 & 1 46 12.4 &-39 23 06 &0.244 &15.88 &15.57& 15.87& 15.56& 15.07&
 \nod &\nod &\nod &-24.88&\nod\nl
 0203-08 & 2 03 22.5 &-8 43 37 &0.770 &16.20 &16.50& 16.90&\nod  & 15.84&
15.48&15.12&-27.03&$<$24.5\nl
 1245-03 &12 47 35.0 &-3 50 09 &0.379 &15.93 &15.04& 16.07& \nod &\nod  &
15.41&14.80&13.77&-25.73&$<$23.8\nl
 1249-02 &12 51 51.4 &-2 23 36 &1.184 &17.00 &16.45& 17.10& \nod &\nod  &
15.80&15.30&14.84&-27.19&25.7\nl
 1249-06 &12 51 56.4 &-7 05 02 &1.187 &16.58 &15.22& \nod & \nod &\nod  &
15.52&14.99&14.55&-27.61&\nod\nl
 1252+02 &12 55 19.6 & 1 44 13  &0.345 &15.52 &14.13& 15.48& \nod &\nod  &
\nod & \nod& \nod&-25.95&25.3\nl
 1321-05 &13 24 14.8 &-6 04 38 &0.732 &16.73 &16.16& \nod & \nod &\nod  &
 15.24&14.72&14.06&-26.38&25.6\nl
 1321+28 &13 21 15.9 & 28 47 19 &0.549 &16.88 &16.15& 16.69& 16.20&\nod  &
15.39&14.81&14.09&-25.57&$<$25.3\nl
 1323+29 &13 23 20.9 & 29 10 07 &0.966 &16.91 &16.17& 16.83& 16.70&\nod  &
\nod  &\nod &\nod &-26.84&$<$24.6\nl
 1326-05 &13 29 28.6 &-5 31 36 &0.580 &15.47 &15.02& 15.59& \nod & \nod &
14.51&13.91&13.10&-27.10&\nod\nl
 1351+01 &13 51 28.4 & 1 03 39 &1.086 &17.68 &16.94& 17.00& 16.50& \nod &
\nod  &\nod &\nod &-26.31&$<$24.6\nl
 1355+02 &13 58 24.0 & 2 13 44 &0.955 &16.61 &16.00& \nod & \nod & \nod &
\nod  &\nod &\nod &-27.12&$<$24.5\nl
 1404+09 &14 04 11.1 & 9 37 38 &0.434 &17.22 &\nod & 17.22& 16.40&\nod  &
\nod  &\nod &\nod &-24.72&25.9\nl
 1406-03 &14 06 10.8 & -3 19 13 &0.860 &17.01 &16.21& \nod & 16.70&\nod  &
15.90 &15.63&14.72&-26.48&$<$24.6\nl
 1415+00 &14 15 49.7 &  0 53 56 &1.042 &17.46 &16.91& \nod & 16.80&\nod  &
\nod  &\nod &\nod &-26.45&$<$24.4\nl
 1415-00 &14 15 28.8 &-0 26 34 &1.148 &17.28 &16.61& \nod & 15.90&\nod  &
15.73 &15.33&14.61&-26.84&$<$24.7\nl
 1424-00 &14 26 58.5 & -0 20 57 &0.632 &16.52 &15.99& \nod & \nod &\nod  &
\nod  &\nod &\nod &-26.25&$<$25.3\nl
 1528+28 &15 28 40.6 & 28 25 30 &0.450 &16.44 &15.79& 16.39& 16.60&\nod  &
15.42 &14.67&14.11&-25.58&23.9\nl
 1630+37 &16 30 18.7 & 37 19 03 &0.960 &17.00 &\nod &\nod  &17.00 &\nod  &
\nod  &\nod &\nod &-26.74&$<$25.7\nl
 1633+39 &16 33 02.1 & 39 24 28 &1.023 &16.00 &15.50&\nod  &15.40 &\nod  &
\nod  &\nod &\nod &-27.88&26.4\nl
 2241-24 &22 44 40.4 &-24 03 02 &1.958 &16.95 &16.06& 16.83& 16.65& 16.31&
15.95&15.45&14.94&-28.41&\nod\nl
 2242-47 &22 45 20.4 &-47 04 50 &0.201 &15.28 &14.22& 14.95& 14.64& 14.13&
\nod  &\nod &\nod &-25.07&\nod\nl
 2313-30 &23 16 37.9 &-30 29 30 &0.257 &15.62 &15.02& 15.15& 14.97& 14.89&
\nod  &\nod &\nod &-25.25&\nod\nl
 2321-29 &23 24 26.1 &-28 54 59 &0.279 &16.44 &15.84& 16.03& 15.80& 15.35&
 14.75&13.93&12.90&-24.61&24.1\nl
 2337-33 &23 39 42.7 &-33 31 21 &0.501 &16.91 &16.09& 17.02& 16.66& 16.62&
16.14&15.42&14.37&-25.36&\nod\nl
 2352-34 &23 55 25.5 &-33 57 57 &0.702 &16.92 &15.49& 16.45& 16.47& 16.28&
\nod &\nod &\nod &-26.14&26.9\nl
 2357-35 & 0 00 01.1 &-35 03 35 &0.508 &16.79 &15.48& 16.76& 16.74& 16.51&
16.22&15.33&14.51&-25.51&\nod\nl
\tablenotetext{a}{log W Hz$^{-1}$ at 1.4GHz}
\enddata
\end{deluxetable}

\clearpage

\def\nod{\nodata}
\voffset=-1in
\hoffset=-0.5in
\begin{deluxetable}{ccccccc}
\tablewidth{35pc}
\tablecaption{ISOPHOT Observations LOG}
\tablehead{
\colhead{Source}           & \colhead{Detector}      &
\colhead{$\lambda$} &
\colhead{Int.time}         & \colhead{Aperture}      &
\colhead{Chopper}          & \colhead{RASTER}
\\
\colhead{} & \colhead{} & \colhead{($\mu m$)} &
\colhead{(s)} & \colhead{($^{\prime\prime}$)}
& \colhead{} & \colhead{}
}
\startdata
 0034-70 & P1   & 7.5  & 180 & 18   & T & \nod\\
         & P1   & 11.5 & 180 & 18   & T & \nod\\
         & P2   & 25.0 & 180 & 23   & T & \nod\\
         & C100 & 60.0 & 230 & 43.5 & T & \nod\\
         & C100 &100.0 & 230 & 43.5 & T & \nod\\
 0049-29 & C100 & 60.0 & 720 & 43.5 & \nod & 3x3\\
         & C200 &160.0 & 960 & 89.4 & \nod & 2x4\\
 0051-39 & P1   & 11.5 & 620 & 18   & R & \nod\\ 
         & P3   & 60.0 & 600 & 52 & R & \nod\\ 
         & C200 &160.0 & 600 & 89.4 & R & \nod\\ 
 0059-30 & P1   & 11.5 & 900 & 18   & R & \nod\\ 
         & P3   & 60.0 & 900 & 52 & R & \nod\\ 
         & C200 &160.0 & 600 & 89.4 & R & \nod\\ 
 0105-26 & P1   & 7.5  & 230 & 18   & T & \nod\\
         & P1   & 11.5 & 230 & 18   & T & \nod\\
         & P2   & 25.0 & 230 & 23   & T & \nod\\
         & C100 & 60.0 & 230 & 43.5 & T & \nod\\
         & C100 &100.0 & 230 & 43.5 & T & \nod\\
 0120-28 & P1   & 11.5 &  900& 18   & R & \nod\\ 
         & P3   & 60.0 &  900& 52   & R & \nod\\ 
         & C200 &160.0 &  400& 89.4 & R & \nod\\ 
 0129-40 & C100 & 60.0 & 720 & 43.5 & \nod & 3x3\\
         & C200 &160.0 & 900 & 89.4 & \nod & 2x4\\
 0144-39 & C100 & 60.0 & 720 & 43.5 & \nod & 3x3\\
         & C200 &160.0 & 920 & 89.4 & \nod & 2x4\\
 0203-08 & P1   & 7.5  & 330 & 18   & T & \nod\\
         & P1   & 11.5 & 330 & 18   & T & \nod\\
         & P2   & 25.0 & 330 & 23   & T & \nod\\
         & C100 & 60.0 & 230 & 43.5 & T & \nod\\
         & C100 &100.0 & 230 & 43.5 & T & \nod\\
 1245-03 & C100 & 60.0 & 520 & 43.5 & \nod & 3x3\\
         & C200 &160.0 & 960 & 89.4 & \nod & 2x4\\
 1249-02 & P3   & 60.0 & 480 & 52   & R & \nod\\ 
         & C200 &160.0 & 600 & 89.4 & R & \nod\\ 
 1249-06 & P3   & 60.0 & 330 & 52   & R & \nod\\ 
         & C200 &160.0 & 360 & 89.4 & R & \nod\\ 
 1252+02 & C100 & 60.0 & 680 & 43.5 & \nod & 3x3\\
         & C200 &160.0 & 860 & 89.4 & \nod & 2x4\\
 1321-05 & P3   & 60.0 & 480 & 52   & R & \nod\\ 
         & C200 &160.0 & 360 & 89.4 & R & \nod\\ 
 1321+28 & P1   & 7.5  & 200 & 18   & T & \nod\\
         & P1   & 11.5 & 200 & 18   & T & \nod\\
         & P2   & 25.0 & 200 & 23   & T & \nod\\
         & C100 & 60.0 & 230 & 43.5 & T & \nod\\
         & C100 &100.0 & 230 & 43.5 & T & \nod\\
 1323+29 & P1   & 7.5  & 230 & 18   & T & \nod\\
         & P1   & 11.5 & 230 & 18   & T & \nod\\
         & P2   & 25.0 & 230 & 23   & T & \nod\\
         & C100 & 60.0 & 230 & 43.5 & T & \nod\\
         & C100 &100.0 & 230 & 43.5 & T & \nod\\
 1326-05 & P3   & 60.0 & 270 & 52   & R & \nod\\ 
         & C200 &160.0 & 150 & 89.4 & R & \nod\\ 
 1351+01 & P1   & 7.5  & 450 & 18   & T & \nod\\
         & P1   & 11.5 & 450 & 18   & T & \nod\\
         & P2   & 25.0 & 450 & 23   & T & \nod\\
         & C100 & 60.0 & 230 & 43.5 & T & \nod\\
         & C100 &100.0 & 230 & 43.5 & T & \nod\\
 1355+02 & P3   & 60.0 & 330 & 52   & R & \nod\\ 
         & C200 &160.0 & 350 & 89.4 & R & \nod\\ 
 1404+09 & P1   & 7.5  & 180 & 18   & T & \nod\\
         & P1   & 11.5 & 180 & 18   & T & \nod\\
         & P2   & 25.0 & 180 & 23   & T & \nod\\
         & C100 & 60.0 & 230 & 43.5 & T & \nod\\
         & C100 &100.0 & 230 & 43.5 & T & \nod\\
 1406-03 & P1   & 7.5  & 450 & 18   & T & \nod\\
         & P1   & 11.5 & 450 & 18   & T & \nod\\
         & P2   & 25.0 & 450 & 23   & T & \nod\\
         & C100 & 60.0 & 230 & 43.5 & T & \nod\\
         & C100 &100.0 & 230 & 43.5 & T & \nod\\
 1415+00 & P1   & 7.5  & 300 & 18   & T & \nod\\
         & P1   & 11.5 & 300 & 18   & T & \nod\\
         & P2   & 25.0 & 300 & 23   & T & \nod\\
         & C100 & 60.0 & 230 & 43.5 & T & \nod\\
         & C100 &100.0 & 230 & 43.5 & T & \nod\\
 1415-00 & P1   & 7.5  & 300 & 18   & T & \nod\\
         & P1   & 11.5 & 300 & 18   & T & \nod\\
         & P2   & 25.0 & 300 & 23   & T & \nod\\
         & C100 & 60.0 & 230 & 43.5 & T & \nod\\
         & C100 &100.0 & 230 & 43.5 & T & \nod\\
 1424-00 & P3   & 60.0 & 330 & 52   & R & \nod\\ 
         & C200 &160.0 & 350 & 89.4 & R & \nod\\ 
 1528+28 & P1   & 7.5  & 330 & 18   & T & \nod\\
         & P1   & 11.5 & 330 & 18   & T & \nod\\
         & P2   & 25.0 & 330 & 23   & T & \nod\\
         & C100 & 60.0 & 230 & 43.5 & T & \nod\\
         & C100 &100.0 & 230 & 43.5 & T & \nod\\
 1630+37 & P1   & 7.5  & 200 & 18   & T & \nod\\
         & P1   & 11.5 & 200 & 18   & T & \nod\\
         & P2   & 25.0 & 200 & 23   & T & \nod\\
         & C100 & 60.0 & 230 & 43.5 & T & \nod\\
         & C100 &100.0 & 230 & 43.5 & T & \nod\\
 1633+39 & P1   & 7.5  & 180 & 18   & T & \nod\\
         & P1   & 11.5 & 180 & 18   & T & \nod\\
         & P2   & 25.0 & 180 & 23   & T & \nod\\
         & C100 & 60.0 & 250 & 43.5 & T & \nod\\
         & C100 &100.0 & 250 & 43.5 & T & \nod\\
 2241-24 & P1   & 11.5 & 900 & 18   & R & \nod\\ 
         & P3   & 60.0 & 900 & 52   & R & \nod\\ 
         & C200 &160.0 & 600 & 89.4 & R & \nod\\ 
 2242-47 & C100 & 60.0 & 720 & 43.5 & \nod & 3x3\\
         & C200 &160.0 & 600 & 89.4 & \nod & 2x4\\
 2313-30 & C100 & 60.0 & 900 & 43.5 & \nod & 3x3\\
         & C200 &160.0 & 720 & 89.4 & \nod & 2x4\\
         & P1   & 11.5 & 330 & 18   & R & \nod\\ 
         & P3   & 60.0 & 330 & 52   & R & \nod\\ 
         & C200 &160.0 & 330 & 89.4 & R & \nod\\ 
 2321-29 & P1   & 11.5 & 900 & 18   & R & \nod\\ 
         & P3   & 60.0 & 900 & 52   & R & \nod\\ 
         & C200 &160.0 & 600 & 89.4 & R & \nod\\ 
 2337-33 & P1   & 11.5 & 900 & 18   & R & \nod\\ 
         & P3   & 60.0 & 900 & 52   & R & \nod\\ 
         & C200 &160.0 & 600 & 89.4 & R & \nod\\ 
 2352-34 & P1   & 11.5 & 900 & 18   & R & \nod\\ 
         & P3   & 60.0 & 900 & 52   & R & \nod\\ 
         & C200 &160.0 & 600 & 89.4 & R & \nod\\ 
 2357-35 & P1   & 11.5 & 900 & 18   & R & \nod\\ 
         & P3   & 60.0 & 900 & 52   & R & \nod\\ 
         & C200 &160.0 & 600 & 89.4 & R & \nod\\ 
\enddata
\end{deluxetable}

\clearpage
\def\nod{\nodata}
\voffset=-1.2in
\hoffset=-0.8in
\begin{deluxetable}{rcccccccccc}
\tablewidth{46pc}
\tablecaption{far-IR Photometry of Quasars$^a$}
\tablehead{
\colhead{Name}           & \colhead{F$_{7\mu m}$}      &
\colhead{F$_{12\mu m}$}          & \colhead{F$_{25\mu m}$}  &
\colhead{F$_{60\mu m}$}          & \colhead{F$_{100\mu m}$}  &
\colhead{F$_{160\mu m}$}         & \colhead{F$_{\rm IRAS 12}$}  &
\colhead{F$_{\rm IRAS 25}$}      & \colhead{F$_{\rm IRAS 60}$}  &
\colhead{F$_{\rm IRAS 100}$}
}
\startdata
 0034-70 & $<$60& $<$60 &$<$90  & $<$160  & $<$130 & \nod  & $<$60 & $<$50 & $<$80 & $<$210\\
 0049-29 & \nod & \nod  & \nod  & $<$105  & \nod   & $<$66 & $<$60 & $<$75 & $<$120& $<$300\\
 0051-39 & \nod & $<$90 & \nod  &36$\pm$18& \nod   &438$\pm$84&$<$120&$<$120&$<$150& $<$450\\
 0059-30 & \nod & $<$30 & \nod  & $<$41   & \nod   &225$\pm$74&$<$90&$<$120&$<$105& $<$360\\
 0105-26 &$<$100& $<$180&$<$210 & $<$100  &$<$140  &\nod   &$<$210&$<$120&{\bf 170$\pm$40}&$<$300\\
 0120-28 & \nod & $<$30 & \nod  &104$\pm$60&\nod   &$<$300 & $<$90&$<$90&{\bf
180$\pm$40}&{\bf 240$\pm$60}\\
 0129-40 & \nod & \nod  & \nod  &100$\pm$30& \nod  &$<$90  & $<$90 &$<$120 &$<$135 & $<$300\\
 0144-39 & \nod & \nod  & \nod  & 74$\pm$20& \nod &76$\pm$26&$<$75 &$<$120 &{\bf110$\pm$30}&{\bf 270$\pm$70}\\
 0203-08 & $<$30& $<$60 &$<$100 &$<$110   &$<$100 &\nod    & $<$60 &$<$150 &$<$150 &$<$400\\
 1245-03 & \nod & \nod  & \nod  &87$\pm$22& \nod  &50$\pm$35&$<$90&$<$240 &$<$135 &$<$240\\
 1249-02 & \nod & \nod  & \nod  &75$\pm$68& \nod   & $<$90  &$<$120&$<$200&$<$120&$<$300\\
 1249-06 & \nod & \nod  & \nod  &217$\pm$72&\nod   &$<$150&$<$120&$<$150&$<$100&$<$300\\
 1252+02 & \nod & \nod  & \nod  &62$\pm$25& \nod   &76$\pm$20
 &{\bf 110$\pm$30}&{\bf 150$\pm$50}&$<$90&$<$400\\
 1321-05 & \nod & \nod  & \nod  &160$\pm$50& \nod  &$<$150  &$<$100 &$<$220 &$<$150 &$<$220\\
 1321+28 & $<$300&$<$300 &60$\pm$25&280$\pm$130&290$\pm$95&\nod&$<$90 &{\bf 120$\pm$30}&$<$110 &$<$420\\
 1323+29 & $<$300&$<$90  &$<$300 &$<$110   &$<$150  &\nod &$<$120 &$<$90 &$<$150 &$<$300\\
 1326-05 & \nod & \nod  & \nod  &160$\pm$80& \nod  &$<$260 &$<$90 &$<$210&{\bf 100$\pm$30}&{\bf 260$\pm$70}\\
 1351+01 & $<$45&$<$90  &60$\pm$20&160$\pm$60&370$\pm$130& \nod &$<$90&{\bf 120$\pm$30}&$<$90 &$<$360\\
 1355+02 & \nod & \nod  & \nod  &264$\pm$78&\nod   &$<$138&{\bf 130$\pm$40}&$<$105&$<$80 &{\bf 200$\pm$70}\\
 1404+09 & $<$90& $<$270 &70$\pm$25&380$\pm$140&566$\pm$160 &\nod &$<$120 &$<$105 &$<$120 &$<$360\\
 1406-03 & $<$45& $<$40 & $<$180 &310$\pm$130&546$\pm$110&\nod &$<$90 &$<$150 &$<$160 &$<$600\\
 1415+00 & $<$45& $<$60 &54$\pm$16&790$\pm$128&280$\pm$80&\nod &$<$135&{\bf
180$\pm$40}&$<$120&{\bf 390$\pm$100}\\
 1415-00 & $<$130&$<$60 &56$\pm$20&475$\pm$120&190$\pm$120&\nod &$<$99
&$<$150&{\bf 140$\pm$40}&{\bf 280$\pm$100}\\
 1424-00 & \nod & \nod  & \nod  &69$\pm$35& \nod   &$<$120 & $<$120 &$<$120 &$<$135 &$<$210\\
 1528+28 & $<$60&$<$30  &50$\pm$40&180$\pm$140&$<$120&\nod &$<$80&$<$60&$<$120&$<$220\\
 1630+37 & $<$45& $<$54 &63$\pm$20&$<$180 &150$\pm$120 &\nod & $<$60  &$<$ 70 &$<$ 90 &$<$270\\
 1633+39 & $<$200& $<$240 &$<$120 &$<$100   &$<$100 &\nod &$<$60 &$<$75 &$<$ 75 &$<$120\\
 2241-24 & \nod & $<$24 & \nod  &45$\pm$23& \nod & 141$\pm$83  &$<$ 90 &$<$150 &$<$90 &$<$240\\
 2242-47 & \nod & \nod  & \nod  & $<$62   & \nod & $<$40 &$<$90 &$<$60 &$<$120 &$<$360\\
 2313-30 & \nod & $<$30 & \nod  &77$\pm$20& \nod & $<$50 &{\bf 90$\pm$20}&$<$150
&{\bf 130$\pm$30}&$<$600\\
         & \nod &       &       &98$\pm$84&      & $<$100& & & &\\
 2321-29 & \nod & $<$20 & \nod  &110$\pm$50&\nod & 280$\pm$80   &\nod & \nod & \nod & \nod \\
 2337-33 & \nod & $<$27 & \nod  & $<$36   & \nod & $<$80 &\nod  &\nod & \nod & \nod \\
 2352-34 & \nod & $<$18 & \nod  & $<$33   & \nod & $<$76 &$<$90 &$<$90 & $<$100 & $<$300\\
 2357-35 & \nod & $<$30 & \nod  & $<$30   & \nod &290$\pm$72&{\bf 140$\pm$30}&
$<$120&{\bf 140$\pm$30} &{\bf 230$\pm$70}\\
\tablenotetext{a}{Fluxes are in mJy}
\enddata
\end{deluxetable}

\end{document}